\documentclass[12pt]{article}
\usepackage{epsfig}
\usepackage{graphicx}
\usepackage{cite}
\usepackage{amsfonts}
\usepackage{amssymb}
\usepackage{latexsym}
\def\ee{\end{equation}}
\def\ba{\begin{eqnarray}}
\def\ea{\end{eqnarray}}
\def\la{~\mbox{\raisebox{-.6ex}{$\stackrel{<}{\sim}$}}~}
\def\ga{~\mbox{\raisebox{-.6ex}{$\stackrel{>}{\sim}$}}~}
\def\bq{\begin{quote}}
\def\eq{\end{quote}}

 at 10truept

\newcommand{\beq}{\begin{equation}}
\newcommand{\eeq}{\end{equation}}
\newcommand{\beqa}{\begin{eqnarray}}
\newcommand{\eeqa}{\end{eqnarray}}
\newcommand{\bea}{\begin{eqnarray}}
\newcommand{\eea}{\end{eqnarray}}
\newcommand{\p}{\partial}
\newcommand{\al}{\alpha}
 \newcommand{\be}{\beta}
 \newcommand{\ep}{\epsilon}

\def\la{~\mbox{\raisebox{-.6ex}{$\stackrel{<}{\sim}$}}~}
\def\ga{~\mbox{\raisebox{-.6ex}{$\stackrel{>}{\sim}$}}~}

\def\ga{~\mbox{\raisebox{-.6ex}{$\stackrel{>}{\sim}$}}~}
\def\ltap{\ \raise.3ex\hbox{$<$\kern-.75em\lower1ex\hbox{$\sim$}}\ }
\def\gtap{\ \raise.3ex\hbox{$>$\kern-.75em\lower1ex\hbox{$\sim$}}\ }
\def\gl{\ \raise.5ex\hbox{$>$}\kern-.8em\lower.5ex\hbox{$<$}\ }
\def\roughly#1{\raise.3ex\hbox{$#1$\kern-.75em\lower1ex\hbox{$\sim$}}}

\def\eps{\epsilon}

\renewcommand{\thefootnote}{\fnsymbol{footnote}}

\begin{document}

\thispagestyle{empty}
\begin{flushright}
BRX-TH 6301\\
November 2015
\end{flushright}
\vspace*{.05cm}
\begin{center}
{\Large \bf Large Field Inflation and Gravitational Entropy}

\vspace*{.75cm} {\large Nemanja Kaloper$^{a, }$\footnote{\tt
kaloper@physics.ucdavis.edu}, Matthew Kleban$^{b, }$\footnote{\tt
kleban@nyu.edu},}\\
\vspace*{.3cm} {\large Albion Lawrence$^{c, }$\footnote{\tt
albion@brandeis.edu} and Martin S. Sloth$^{d, }$\footnote{\tt
sloth@cp3.dias.sdu.dk} }\\
\vspace{.5cm} {\em $^a$Department of Physics, University of
California, Davis, CA 95616, USA}\\
\vspace{.3cm} {\em $^b$CCPP, Department of Physics, New York University, NY 10003, USA}\\
\vspace{.3cm} {\em $^c$Martin Fisher School of Physics, Brandeis University, Waltham, MA 02453, USA}\\
\vspace{.3cm} {\em $^d$$CP^3$-Origins, Center for Cosmology and Particle Physics Phenomenology,}
\vspace{.1cm} {\em University of Southern Denmark, Campusvej 55, 5230 Odense M, Denmark}\\

\vspace{.5cm} {\bf Abstract}
\end{center}

Large field inflation can be sensitive to perturbative and nonperturbative quantum corrections that spoil slow roll.
A large number $N$ of light species in the theory, which occur in many string constructions, can amplify these problems. One might even worry that in a de Sitter background, 
 light species will lead to a violation of the covariant entropy bound at large $N$. If so, requiring the validity of the covariant entropy bound could limit the number
of light species and their couplings, which in turn could severely constrain axion-driven inflation. Here we show that there is no such problem when 
we correctly renormalize  models with many light species, taking the {\it physical} Planck scale to be $M^2_{pl} \ga N {\cal M}_{UV}^2$, where ${\cal M}_{UV}$ is the cutoff for the QFT coupled to semiclassical quantum gravity. The number of light species then cancels out of the gravitational entropy of de Sitter or near-de Sitter backgrounds at leading order.
Working in detail with $N$ scalar fields in de Sitter space, renormalized to one loop order, we show that the gravitational entropy automatically obeys the covariant entropy bound. Furthermore, while the axion decay constant is a strong coupling scale for the axion dynamics, we show that it is {\it not} in general the cutoff of 4d semiclassical gravity. After renormalizing the two point function of the inflaton, we note that it is also controlled by scales much below the cutoff.  We revisit $N$-flation and KKLT-type compactifications in this light, and show that they are perfectly consistent with the covariant entropy bound. Thus, while quantum gravity might yet spoil large field inflation, holographic considerations in the semiclassical theory do not obstruct it.

\vfill \setcounter{page}{0} \setcounter{footnote}{0}
\newpage

\renewcommand{\thefootnote}{\arabic{footnote}}
\setcounter{equation}{0} \setcounter{footnote}{0}

\section{Introduction}

Inflation is the best theoretical explanation of the large, old and smooth universe with small nearly scale invariant perturbations. It fits experimental tests perfectly. If, in addition to the observed scalar density fluctuation, tensor mode fluctuations of the CMB are directly observed,
we could probe inflation in great detail. If these modes are generated by quantum fluctuations of the gravitational field, a direct observation by itself would imply that the effective field theory (EFT) models of inflation required to generate large tensor modes would have to operate up to scales around the GUT scale, which is close to the string or 10d/11d Planck scale in many conservative string theory scenarios. Moreover, to yield sufficiently long inflation such models would have to have very flat potentials over super-Planckian field ranges \cite{Lyth:1996im,Efstathiou:2005tq}. 

At such scales and over such field ranges, quantum field theory and quantum gravity effects correcting the dynamics can be significant. In this light it is reasonable to ask whether quantum gravity might provide any general constraints on these models, independent of a specific realization in string theory. While we do not have a complete theory of inflation in quantum gravity yet, there are hints and clues about how quantum gravity might 
influence the low energy theory. The two lines of inquiry which have received particular attention recently are the ``weak gravity conjecture" (WGC) of \cite{ArkaniHamed:2006dz} and the covariant entropy bounds \cite{Bousso:1999xy}. The WGC uses some features of 
black hole entropy to place an upper bound on the mass of charged particles and/or the action of instantons leading to corrections to the axion potential. This can constrain some axion inflation models such as \cite{Freese:1990rb,ArkaniHamed:2003mz,ArkaniHamed:2003wu,Dimopoulos:2005ac}\ for which the axion potential is of the sinusoidal form expected from the dilute gas approximation for instantons.  There are more tenuous arguments \cite{Brown:2015iha,Heidenreich:2015wga,Palti:2015xra}\ that axion monodromy inflation \cite{Silverstein:2008sg,McAllister:2008hb,Kaloper:2008fb,Kaloper:2011jz,DAmico:2012ji,DAmico:2012sz,DAmico:2013iaa,Palti:2014kza,Hebecker:2014eua,Ibanez:2014kia,Ibanez:2014swa,Bielleman:2015lka,Bielleman:2015ina,Grimm:2014vva,Blumenhagen:2014gta,Blumenhagen:2014nba,Marchesano:2014mla,Gao:2014uha}\ is constrained by such considerations as well, although at present they are not excluded.

The argument from the covariant entropy bound \cite{Conlon:2012tz}\ is based on the following logic. If there are  many light weakly interacting species and a sufficiently long inflationary epoch\footnote{For earlier considerations of bounds on duration of inflation see \cite{Banks:2003pt,Kaloper:2004gp}.} (or, a long lived metastable de Sitter space), the large number of species will thermalize with gravity and overwhelm the geometric entropy. 
The setup is especially interesting as string-motivated inflationary constructions 
often consist of low energy theories with $N$ light species, $N \gg 1$, which are weakly coupled to each other in the IR. According to the argument above, such models violate the covariant entropy bound, which requires that the entropy in a given Hubble-sized patch does not exceed the Gibbons-Hawking entropy. Requiring the validity of the covariant entropy bound thus constrains $N$, the couplings of these species, and the duration of inflation (or the lifetime of the metastable de Sitter space). 

In what follows we will focus on this latter issue\footnote{We stress that while this issue and its analysis at present appear distinct from the arguments about WGC in \cite{ArkaniHamed:2006dz}, ref. \cite{Brown:2015iha}\ has suggestion a strong form of the WGC may be related to the entropic arguments in \cite{Conlon:2012tz}.  We have nothing to say about this possibility, but it is true that both are ultimately related to entropic considerations.}. We specifically show that even for field theories with many light species, a correctly renormalized low energy theory, including the gravitational couplings, obeys the covariant entropy bound since the number of light species cancels from the entropy formulas to the leading order. Essentially, this arises since even if one starts with many weakly coupled species of particles, when gravity is turned on the renormalized effective field theory becomes strongly coupled well below the Planck scale, ${\cal M}_{UV} \la M_{pl}/\sqrt{N}$, which sets a cutoff on calculations using weakly coupled, semiclassical gravity. Beyond this scale, one needs a full-blown UV completion to follow the details of the dynamics. Nevertheless, the low energy description remains self-consistent, and obeys the semiclassical limits, as one might have expected from decoupling.

In more detail, the argument that entropy bounds  are violated begins with an estimate of the contribution to the total de Sitter entropy of a field $\phi$. The first step is the identification of a cutofff ${\cal M}_{UV}$ for the dynamics of $\phi$. For axions, this is identified with the axion decay constant $f$, which sets the periodicity in field space $\phi \equiv \phi + 2\pi f$. The next step is to count the number of patches of size ${\cal M}_{UV}$ on the de Sitter horizon, leading to a contribution $S_{\phi} \sim  {\cal M}_{UV}^2/H^2$, where $H$ is the Hubble scale.  One then demands that $\sum_i S_{\phi_i} \leq \frac{M_{pl,4}^2}{H^2}$.  Then if $\sqrt{N}  {\cal M}_{UV} > M_{pl,4}$, there appears to be an apparent violation of the covariant entropy bound due to a species problem. 

The hole in this argument is that the formula $S \sim A/(4 G_N)$ for 4d de Sitter or inflating universes is valid only when the underlying (semiclassical) 4d gravity is valid. This means that one must take into account the loop corrections to the gravitational sector, and consistently analyze the {\it renormalized} 4d effective field theory of gravity, accounting for the contributions of all the many light species 
to the relevant physical quantities which control the dynamics, including the cutoff and the dimensional couplings. In particular 
the route to the proper renormalization procedure must incorporate the following:
\begin{itemize}
\item The correct cutoff to impose is the scale at which 4d semiclassical gravity breaks down. This is generically at a scale $ {\cal M}_{UV}^2 \leq M_{pl}^2/N$. This cutoff follows from the well known behavior of perturbative renormalization of gravity which involves the inclusion of higher dimension of irrelevant operators in the gravitational sector, which introduce a perturbative spin-2 ghost, with a mass $\sim 1/ {\cal M}_{UV}$ \cite{Stelle:1976gc}. 
\item The correct value of $G_N \sim 1/ M_{pl}^2$ to use is the renormalized Newton's constant at this scale, $1/G_{N, ren} \sim M_{pl}^2 \sim M_{pl, bare}^2 + N  {\cal M}_{UV}^2$. The inclusion of the contribution of $N$ species at the scale ${\cal M}_{UV}$ evades the species problem. 
\item The axion decay constant $f$ and the scale ${\cal M}_{UV}$ may be very different. Specifically, ${\cal M}_{UV}$ can be much smaller than the period. This may arise very simply in setups with intermediate-mass particles, as we will show explicitly using a two-axion model \cite{Kim:2004rp,Berg:2009tg}.
\end{itemize}

Footnote 2 in \cite{Conlon:2012tz}\ dismisses significant renormalization of Newton's constant in models with many species of light fields, claiming there can be cancellations between corrections. The calculations in \cite{Kabat:1995eq,Larsen:1995ax}\ show that while minimally coupled scalars and Weyl fermions contribute with the same sign, Abelian gauge fields contribute with the opposite sign.  However, this is not a way out. First, for any field with spin less than 2, the divergent contributions to the entanglement entropy defined via the replica trick \cite{Callan:1994py}\ have been shown to be {\it precisely}\ taken into account by the same fields' contribution to the renormalization of the gravitational action -- see \cite{Jacobson:2012ek,Cooperman:2013iqr}, and the references therein.\footnote{The interpretation of the gauge field contributions to the entanglement entropy is a subject of ongoing research \cite{Donnelly:2011hn,Casini:2013rba,Huang:2014pfa,Ghosh:2015iwa,Solodukhin:2015hma,Donnelly:2015hxa,Soni:2015yga,Ma:2015xes}.}  Secondly, Newton's constant is not the only place that quantum corrections to the gravitational action will occur, and the relative contributions of different fields will differ in these other terms, so that the estimate $M_{pl}/\sqrt{N}$ of the strong coupling scale remains appropriate.

While the itemized points above are individually discussed in the literature, the recurring confusions suggest that a unified and coherent discussion in the context of large field inflation is warranted. 
Hence we will provide a thorough review of these arguments (\S2), some additional calculations supporting them in cosmological settings (\S3), and a re-examination (\S4) of the claimed constraints on $N$-flation and KKLT-type compactifications that are explicitly discussed in \cite{Conlon:2012tz}.

\section{The strong coupling scale for gravity and the ``species problem"}

The classic calculations of the Bekenstein-Hawking entropy for black holes, and the Gibbons-Hawking entropy for cosmological spacetimes, are based crucially on semiclassical gravity. One can only use these results in the regime of validity of the (renormalized) semiclassical theory, and one must use the physical, renormalized couplings and scales at these energies. In this section we will describe the renormalization procedure and extract the behavior of the renormalized quantities, particularly the renormalized Planck scale, as a function of the number of light species of particles which appear in the loops that contribute to the effective action of gravity. We will also note that the higher dimension irrelevant operators in the gravitational sector, which arise from the loop corrections, yield a clear cutoff that determines the validity of the semiclassical approximation. These points have been noted in the context of black holes, via a variety of arguments elucidated below. We will close the section by recalling the emergence of the strong coupling scale in compactifications from $d > 4$ theories, and in the Randall-Sundrum scenario.

\subsection{Perturbation theory arguments}

We will first revisit perturbative renormalization of gravity due to exchange of virtual field theory degrees of freedom, starting with Einstein-Hilbert theory perturbatively quantized around a vacuum with maximal symmetry. The case of the Minkowski vacuum has been studied extensively \cite{tHooft:1974bx,Stelle:1976gc,Adler:1982ri}. The generalization to de Sitter vacua is straightforward, and has been considered in the context of computing black hole entropy in curved backgrounds. Here we will follow \cite{Fursaev:1994ea,Larsen:1995ax,Demers:1995dq}. In this subsection we will focus on the renormalization of the action. Later on we will see how those results affect the horizon entropy.

If we start with the bare gravitational Lagrangian  
\beq
\mathcal{L}_g = \frac{1}{16\pi G_{N}}(R-2\Lambda) - {\cal L}_{m}(\phi) +\frac{1}{4\pi}\left[a R^2 +b R_{\mu\nu}R^{\mu\nu}+c R_{\mu\nu\kappa\rho}R^{\mu\nu\kappa\rho}\right]
\label{action}
\eeq 
where ${\cal L}_{m}$ is the quantum field theory of some matter, the bare quantities are $G_N$, the bare Newtons constant, $\Lambda$, the bare cosmological constant, and $a$, $b$, $c$, the higher dimension irrelevant operator bare couplings. These terms are set to cancel the one loop divergences in the theory due to matter couplings \cite{tHooft:1974bx}. For simplicity we will take the matter sector to consist of $N$ minimally coupled scalar fields, with only quadratic Lagrangians,
\beq
\mathcal{L}_m = \sum_{j=1}^N \frac{1}{2}\left[ \p_{\mu}\phi_j \p^{\mu}\phi_j+m_j^2\phi_j^2\right] \, .
\label{mattlag}
\eeq
This is sufficient for our purposes. Generalizations to other matter sectors are straightforward, and as we will discuss below, do not change the essential conclusions.

The one-loop contributions to the effective action from the matter sector will generically exhibit quartic, quadratic and logarithmic UV divergences \cite{tHooft:1974bx}. The quartic UV divergence is the usual divergent contribution to the cosmological constant. If we truncate the matter theory to the quadratic Lagrangian (\ref{mattlag}) it may or may not appear depending on the regulator. The quadratic divergences are the wave-function renormalizations of the kinetic terms in (\ref{action}), and include renormalizations of the additional ``$R^2$" terms in the action. 
\vskip.5cm
\begin{figure}[thb]
\label{diagram}
\centering
\includegraphics[height=3cm]{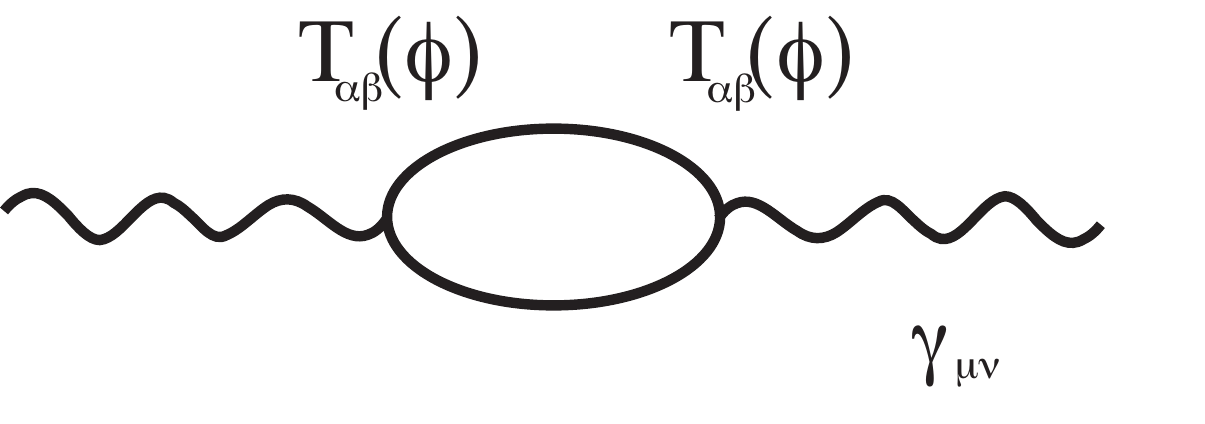}
\caption{One loop graviton vacuum polarization diagram.}
\end{figure}
\vskip.5cm

To compute the one loop integrals, one first needs to regulate the divergent terms. We do so by introducing a  system of Pauli-Villars regulators for every matter field in (\ref{mattlag}). The scheme is conceptually the same as in flat space, where one introduces a regulator for every divergent loop. If the cutoff is above the inverse curvature scale we can start with a locally flat region of space, introduce the regulators with minimal coupling to gravity. Because there are five distinct types of required counterterms, reflecting five different divergences, one needs five regulators for each matter scalar 
\cite{Demers:1995dq}, denoted by $\phi_i$ ($i=1,\dots,5$) and coming with different statistics, $\Delta_i$  (where $\Delta_i =\pm 1$ for commuting and anticommuting fields respectively). The regulator masses $m_i$ are much larger than the matter ones in order to formally cancel the UV divergences, and define the UV cutoff, $\mu$. The choice of the regulators and their statistics are determined by the requirements 
\beq\label{sumrules}
\sum_{i=0}^5 \Delta_i =0~\qquad \textrm{and}\qquad \sum_{i=0}^5 \Delta_i m_i^2=0
\eeq
ensuring finiteness of the regulated theory. Here $m_0$ is the mass of the original scalar field $\phi\equiv \phi_0$. Using this regularization procedure, the one-loop effective action is given by \cite{Demers:1995dq}
\bea\label{LR}
\mathcal{L}_g&=&-\frac{1}{8\pi}\left(\frac{\Lambda}{G_N}+\frac{\gamma}{4\pi}\right)+\frac{R}{16\pi}\left(\frac{1}{G_B}+\frac{\delta}{12\pi}\right)\nonumber\\
&&+\frac{1}{4\pi}\left[\left(a+\frac{\al}{576}\right) R^2 +\left(b-\frac{\al}{1440 \pi}\right) R_{\mu\nu}R^{\mu\nu}+\left(c+\frac{\al}{1440 \pi}\right)  R_{\mu\nu\kappa\rho}R^{\mu\nu\kappa\rho}\right]
\eea
where
\beq
\al = N \sum_{i=0}^5 \Delta_i \log m_i^2~,\qquad \delta = N \sum_{i=0}^5 \Delta_i m_i^2\log m_i^2,\qquad \gamma =  \frac{N}{2}\sum_{i=0}^5 \Delta_i m_i^4\log m_i^2~.
\label{counters}
\eeq
These expressions are in fact dimensionless once the logs are summed up, due to the fact that the $\Delta_i$ are alternating numbers. One thus obtains the renormalized cosmological constant and Newton's constant
\beq
 \frac{\Lambda^{ren}}{G^{ren}_N}= \frac{\Lambda}{G_N}+\frac{\gamma}{4\pi} ~,\qquad \frac{1}{G^{ren}_N} = \frac{1}{G_N} +\frac{\delta}{12\pi}~.
\eeq
Furthermore, the second line of the expression (\ref{LR}) shows how, due to covariantization of the action, the wave function renormalization of the graviton depicted by the Feynman diagram of Fig. (\ref{diagram}) forces the introduction of the counterterms involving the ``$R^2$" terms. 

The renormalized values of Newton's constant and the cosmological constant $G_N^{ren}$ and $\Lambda^{ren}$ are not calculable but are completely arbitrary. They are inputs to the theory which need to be {\it measured}\footnote{See \cite{Kaloper:2013zca,Kaloper:2014dqa,Kaloper:2015jra} for a discussion of the measurement subtleties.}. One needs to put in two renormalization conditions, which specify the values of $G_N^{ren}$ and $\Lambda^{ren}$ at the subtraction point. This implies that the renormalized quantities depend on the subtraction point scale in the same way as the regulator masses. Taking the subtraction point to be the field theory UV cutoff implies that the renormalized Newton's constant depends on it in the same way as on the regulator masses, $1/G_N \sim {\cal O}(1) N {\cal M}_{UV}^2$, since $m_i \sim {\cal M}_{UV}$ for $i = 1, \ldots , 5$. Therefore, the renormalized Planck scale is
\beq
M_{pl}^{2} = {\cal O}(1) N {\cal M}_{UV}^2 + M_{pl, bare}^{2} \,, 
\eeq
where the last term includes any additional contributions from the gravitational sector, additional UV degrees of freedom, and finite IR corrections.

When fermions are included, similar conclusions apply. Indeed, \cite{Larsen:1995ax,Veneziano:2001ah, Dvali:2007wp}\  point out that upon integrating out $N_0$ minimally coupled scalars and $N_{1/2}$ fermions above the cutoff $\Lambda$, the one-loop effective action for the gravitation field has the form
\beq
	S_{1-loop} \sim \int d^4 x \sqrt{g} \left(M_{pl, bare}^2 + c_1 N_0 {\cal M}_{UV}^2 + c_2 N_{1/2} {\cal M}_{UV}^2\right) R \, ,
\eeq
where $c_1, c_2$ have the same sign. The renormalized Planck mass is thus
\beq
	M_{pl}^{2} =  M_{pl,bare}^2 + \left(c_1 N_0 + c_2 N_{1/2}\right) {\cal M}_{UV}^2 \, .
	\label{plbound}
	\eeq
%

In \cite{Kabat:1994vj,Solodukhin:1995ak,Larsen:1995ax}, it was shown that gauge fields and nonminimally coupled scalars contribute negative shifts to the renormalization of Newton's constant. Thus, one may object that in some theories Newton's constant may receive a small renormalization due to cancellations between different fields running in the loops, and our resolution of the species problem does not apply. However, if we compute the entanglement entropy via the replica trick \cite{Callan:1994py}, the divergences in that calculation are nonetheless precisely taken care of by the renormalization of the gravitational action: see \cite{Jacobson:2012ek,Cooperman:2013iqr,Solodukhin:2015hma} and the references therein. The intepretation of the contribution of gauge fields is an active subject of research \cite{Donnelly:2011hn,Casini:2013rba,Ghosh:2015iwa,Solodukhin:2015hma,Donnelly:2015hxa,Soni:2015yga}. Nonetheless, it appears to be consistent to take the entanglement entropy as computed by the replica trick to account for the contribution of light fields to the gravitational entropy \cite{Jacobson:2012ek,Cooperman:2013iqr}. If we do so there is still no species problem.

Furthermore, Newton's constant is not the only term in the gravitational action to get renormalized: the loop contributions to the $(curvature)^2$ will involve different combinations of the effects of different species, without cancellations.

The upshot of this is that any truncated effective theory of QFT coupled to gravity, with higher dimension irrelevant operators constructed from geometric invariants, is strongly coupled beyond ${\cal M}_{UV}$. The action (\ref{LR}) already shows this, since it contains at least a spin-2 massive ghost, with a mass $m_{ghost} \simeq M_{pl}/\sqrt{c_{ren}} \simeq M_{pl}/\sqrt{N}$, where\footnote{Here we ignore the numerical factors in the renormalized value of $c$ because they can be compensated by the logs in realistic models with very light particles.}
$c_{ren} = c+\frac{\alpha}{1440\pi}$ as in (\ref{LR}) \cite{Stelle:1976gc}. The ghost will generically remain present at any finite loop order of the expansion, and without a full UV completion it is impossible to determine if the theory can be extended above this scale. This has been noted previously in the cosmological context in \cite{Simon:1990ic,Simon:1990jn,Simon:1991bm}. 

In summary, to consistently do semiclassical 4d gravity calculations,
one must restrict the theory to scales below the physical cutoff
\beq
{\cal M}_{UV} \la \frac{M_{pl}}{\sqrt{N}}.
\label{cutoffs}
\eeq
 If $M_{pl}$ is fixed by classical gravitational measurements and $N$ is increased, the cutoff of the QFT {\it must} in general be correspondingly lowered. The inequality $M_{pl}^2 \geq \left(c_1 N_0 + c_2 N_{1/2}\right) {\cal M}_{UV}^2$, implied by (\ref{plbound}), will be saturated when $N \gg 1$, yielding the scaling $M_{pl}^{2} = N {\cal M}_{UV}^2$.  This occurs, for example, in RS2 braneworlds \cite{Randall:1999vf,Gubser:1999vj} and in induced gravity \cite{Sakharov:1975zg,Adler:1982ri}.

\subsection{Black hole entropy}

The species problem appears in considerations of black hole entropy when one tries to include the effects of large numbers of matter fields. There are two apparently different approaches which however yield the same answer (see \cite{Jacobson:2012ek,Cooperman:2013iqr}\ for up-to-date versions of the relevant arguments, and for surveys of prior work). 

One approach is to compute the free energy for a black hole as a function of the temperature, via computing fluctuations about the Euclidean saddle point, take the appropriate derivatives with respect to temperature. This gives a thermal entropy whose classical contribution is the Hawking-Bekenstein entropy. 

The other approach is to interpret the black hole entropy as an entanglement entropy. Then the one-loop contributions from the matter fields can be computed following the prescription of \cite{Susskind:1993ws,Callan:1994py,Solodukhin:1994yz}. Technically, they involve changing the temperature without changing the horizon radius, introducing a conical deficit angle into the spacetime. However, the result for the entropy is the same as the saddle point approach given above. In four dimensions, the calculation involves divergences which are quadratic in the cutoff and which scale with the number of species.
Such divergences are absorbed precisely by the renormalization of the gravitational effective action. The resulting entropy will be the Hawking-Bekenstein or Wald entropy for the black hole, with the renormalized Newton's constant that is the correct physical gravitational coupling at low energies.  The species problem never appears so long as one writes the Bekenstein-Hawking entropy in terms of physical couplings. The calculation also confirms that the appropriate cutoff is precisely the formula (\ref{cutoffs}) \cite{Dvali:2008jb}. 

The arguments reviewed above agree with the following qualitative picture of the black hole from the point of view of a static Schwarzschild observer.  Static observers a proper distance $\epsilon$ from the black hole see a local Unruh temperature of order $T_u = 1/\epsilon$.  One can consider the region within a distance $\epsilon$ from the horizon to be a thermal membrane or "stretched horizon" \cite{Susskind:1993if}\ with temperature $T_u$. For N species lighter than $T_u$, the thermal entropy of this membrane is $S = N T_u^3 V$ where $V = \epsilon A$ is the volume of the stretched horizon, $A$ is the area of the black hole.  Setting $\epsilon = {\cal M}_{UV}$, we find $S = N {\cal M}_{UV}^2 A$ which parametrically matches the Hawking-Bekenstein entropy when ${\cal M}_{UV} = M_4/\sqrt{N}$ \cite{Jacobson:1994iw,Hawking:2000da,Emparan:2006ni,Kaloper:2012hu}.  Increasing the cutoff further takes us out of the range of 4d semiclassical gravity, and requires knowledge of the UV completion.

The precise details of the analogous calculations for the gravitational entropy of cosmological horizons are still not available, although there are arguments that the considerations above should extend to such cases \cite{Kim:1998zs,Hawking:2000da}. Furthermore, the qualitative discussion of the paragraph above applies directly to the de Sitter horizon as seen by a static observer.

Since the work of \cite{Strominger:1996sh}, it has been clear that the Bekenstein-Hawking entropy for black holes in string theory should be considered as a statistical entropy, counting the density of states as a function of the energy.  Recent advances in our understanding of the emergence of thermodynamics in closed quantum systems \cite{PhysRevA.43.2046,PhysRevE.50.888,PhysRevLett.80.1373,rigol2008thermalization}\ have shown that the entropy of a thermal system can be understood precisely as an entanglement entropy between observed and unobserved factors of the Hilbert space.  It would be interesting to explore the equivalence between the semiclassical gravity calculations of each, in this light.  It may also help shed light on the nature of de Sitter entropy.

\subsection{Some examples of the strong coupling scale}

We close this section by noting previous concrete examples of a strong coupling scale for gravity below the Planck scale. A very simple example \cite{Dvali:2007hz,Dvali:2007wp}\ is a D-dimensional theory (say, string theory) with a fundamental Planck scale $M_D < M_4$, compactified on a $d = D-4$-dimensional manifold with volume $V = L_{KK}^d$.  The number of KK modes between $L_{KK}$ and $M_D$ is, roughly, $(L_{KK} M_D)^d$; the strong coupling scale is then
\beq
	{\cal M}_{UV}^2 = \frac{M_4^2}{(L_{KK} M_D)^d} = \frac{L_{KK}^d M_{KK}^{d-2}}{(L_{KK} M_D)^d} = M_D^2
\eeq
so 4d semiclassical gravity breaks down at the fundamental $D$-dimensional Planck scale $M_D$.

We can also consider a general case in which $N_3$ D3-branes are close together within some 6d compactification with volume $V \sim R_{KK}^6$; this corresponds to the UV completion of the previously mentioned RS braneworld scenario \cite{Randall:1999ee,Randall:1999vf}.  In this case,  $N_3^2 = (R_{ads}/\ell_{p,10})^8$ is the number of light species, where $R_{ads}$ is the scale of warping near the D3-branes; deep in the core of the D3-brane geometry, it is the curvature radius of the associated anti-de Sitter background.  The strong coupling scale for 4d gravity is
\beq
	{\cal M}_{UV}^2 = \frac{m_{pl}^2}{N_3^2} \sim \frac{R_{KK}^6}{R_{ads}^8} \, .
\eeq
Now, if $R_{ads} < R_{KK}$, so that the D3-brane throat is smaller than the KK scale, ${\cal M}_{UV}^2 \gg 1/R_{ads}^2$, and the theory becomes effectively five-dimensional at scales below ${\cal M}_{UV}^2$.  As we increase the number of D3-branes, we have a strongly warped compactification that is well-described by a RS braneworld scenario \cite{Verlinde:1999fy,Gubser:1999vj}. The 4d Planck scale is $\delta m_{pl}^2 = m_{pl,5}^3 R_{ads}$, where $m_{pl,5}$ is the 5d Planck scale; the central charge of the dual CFT is $c\propto (R m_{pl,5})^3$. This implies a UV scale ${\cal M}_{UV} \sim \frac{m_{pl}}{\sqrt{c}} \sim 1/R_{ads}$. Thus the theory becomes effectively five-dimensional at ${\cal M}_{UV}$.

\section{Renormalization of de Sitter entropy}

To determine the gravitational entropy of $N$ massive (light) scalars in de Sitter space we work in the "static patch" of $3+1$-dimensional de Sitter space, with the metric
\beq \label{static}
ds^2 = -g(r)d\tau^2+\frac{1}{g(r)}dr^2+r^2d\Omega^2_3~,\qquad
g(r)=\left(1-\frac{r^2}{r_H^2}\right)~,
\eeq
where $H^2 = 1/r_H^2$ is the de Sitter Hubble constant. The section of the geometry $r \le r_H$ is the causal patch of a single observer at $r=0$, and  $r_H$ is the location of her event horizon. One can define a Hamiltonian as the infinitesimal generator of translations in static time $\tau$.  Let $S(\beta)$ be the thermal entropy at the Gibbons-Hawking temperature 
$T = 1/\beta = H/2\pi$ computed in the canonical ensemble defined with respect to this static patch Hamiltonian. The covariant entropy bound states that $S(r_H) = A/4 G_N$. 

We want to determine the contribution of $N$ scalars to this entropy\footnote{A similar calculation was done in \cite{Kim:1998zs}\ for $2+1$-dimensional de Sitter space.}.  The blueshift near the horizon implies that a large number of modes are concentrated there, and leads to a divergence.  To deal with these and relate them to the renormalization of the gravitational effective action, we follow the strategy of \cite{Demers:1995dq} which ensures that regularization of the entropy and the gravitational effective action are done in the same scheme \cite{Kabat:1994vj}. First, we impose ``brick wall"  boundary conditions (ie Dirichlet boundary conditions) on all scalar fields at a small but finite distance from the horizon. This surface acts not only as a position-space regulator,  but also a momentum space regulator, by isolating the leading quadratic divergence coded by the blueshift in the near horizon limit. We use the renormalization prescription for the one-loop gravitational effective action discussed above to renormalize the entropy. The resulting contribution to the gravitational entropy precisely matches the renormalization of Newton's constant, extending the Bekenstein-Gibbons-Hawking formula to one loop in de Sitter space. We close with some comments regarding the relationship to entanglement entropy.

\subsection{Renormalizing the entropy of $N$ scalar fields}

Consider a free, massive scalar field in de Sitter space. 
To regulate the theory in this background, we impose the ``brick-wall" boundary condition
 \beq
\Phi = 0 \qquad\textrm{at}\qquad  r=r_H-\ep~.
 \eeq
Here $\ep$ is the coordinate distance of the brick wall regulator from the horizon $r_H = 1/H$. An an infrared cutoff is not necessary since the static patch of de Sitter is finite. We now compute the free energy of this scalar at a temperature $T$, which in the end we will set to be the Gibbons-Hawking temperature $T_{GH} = H/2\pi$.

We next determine the mode expansion for the the energy levels $E(n,l,l_3)$ of the field $\Phi$. Here $l$ is the total angular momentum and $l_3$ the angular momentum along some fixed axis. The field equation for the
modes with energy $E$ and angular momentum quantum numbers $l,l_3$ are $\Phi = e^{-iE\tau} Y_{l}^{l_3}(\theta,\varphi) \phi(r)$, where the radial modes obey 
  \beq
\frac{1}{r^2}\p_r
\left(r^2g(r)\p_r\phi \right) + \left(\frac{1}{g(r)}E^2- m^2 - \frac{l(l+1)}{r^2}\right)\phi=0~.
  \eeq
Close to the horizon, the energy blueshift as $g(r) \rightarrow 0$ guarantees that the WKB approximation $\exp(\pm i\int k(r)dr)$
where
 \beq \label{kmom}
k^2(r,l,E)=
\frac{1}{g^2(r)}E^2 - \frac{1}{g(r)}\left(m^2 + \frac{l(l+1)}{r^2}\right)~ \, ,
 \eeq
will give a good accounting of the behavior and multiplicity of modes with a given energy, $\pi n= \int_{L}^{r_H-\ep}dr k(r,l,E)$. Moreover, the blueshift also guarantees that this region gives the dominant contribution to the entropy which diverges in the near horizon limit. Hence, the leading order contributions to the entropy will come from precisely the modes in this regime. Within this approximation, the number of states up to energy $E$ is
\beq \label{rho}
	\rho(E) = {1 \over \pi} \int_0^{r_H - \eps} dr \int_0^{l_{max}(E)} dl (2l + 1) \frac{\sqrt{E^2 - g(r) (m^2 + \frac{l(l+1)}{r^2})}}{g(r)} \, ,
\eeq
where $l_{max}$ is the value at which the argument of the square root vanishes, and the summation of angular momenta $l$ is approximated by an integral, which is valid for $l \gg 1$ near the horizon.

Now, for $N$ identical scalars, the free energy at inverse temperature $\be$ is given by
 \beq \label{free1}
e^{-\be F}=\prod_{n,l,l_3}\frac{1}{(1-e^{-\be E(n,l,l_3)})^N}~.
 \eeq
Hence,  $\beta F = N \int dE \, \frac{d\rho(E)}{dE} \ln(1-\exp(-\beta E)) = - \beta N \int dE \, \rho(E)/(e^{\beta E} - 1)$ after integration by parts.
Further, following \cite{tHooft:1984re}, since the dominant contributions will come from the highest energy modes, which have large $l$, their density of states and total number of modes with a given energy behave as $l(l+1) \sim l^2$, $(2l + 1) \sim 2l$. Integrating over $l$ in (\ref{rho}) then gives
 \beq \label{free2}
F = -\frac{2N}{3\pi }\int_0^\infty dE\frac{1}{e^{\be
E}-1}\int^{r_H-\ep}_{0}dr\frac{r^2}{g^2(r)}\left(E^2-g(r)m^2\right)^{3/2} \, .
 \eeq
Rewriting $g(r) = - \frac{2}{r_H} (r-r_H) - \frac{1}{r_H^2}(r-r_H)^2$ to extract the divergences
in the limit $\ep \to 0$ we find
 \bea \label{free3}
 F &=& -\frac{2 N}{3\pi}\int_0^\infty dE\frac{ E^3}{e^{\be
E}-1}\frac{r_H^4}{4}\left[\frac{1}{\ep} +\left(\frac{1}{r_H} + \frac{3 m^2}{r_H E^2} \right)\log(\epsilon/r_H)+\mathcal{O}(1)\right]\nonumber\\
&=&
-\frac{N \pi^3}{90}\frac{r_H^4}{\be^4} \frac{1}{\ep} 
-\frac{N \pi^3}{90}\frac{r_H^4}{\be^4}  \left(\frac{1}{r_H} + \frac{15}{2}\frac{\be^2 m^2}{\pi^2 r_H} \right)\log(\epsilon/r_H)+ {\rm finite ~ terms} ~\, .
 \eea

The free energy has two divergences, an inverse power and a logarithmic one. These are the cutoff-dependent contributions which are subtracted off by the counterterms of the theory, defined by the regulators. 
In order to subtract these divergences using the prescription for the renormalization of the effective action (\ref{free3}), we must compute these quantities in the same scheme \cite{Kabat:1994vj}. 
Since -- as \cite{Demers:1995dq} -- we are using the Pauli-Villars regulators, the total free energy for $N$ scalar fields and the system of Pauli-Villars regulators for each of them is
\beq
\be F = \be \sum_{i=0}^5 \Delta_i F^i
\label{totalfree}
\eeq
where $F^i$ is (\ref{free3}) computed using the mass, $m_i$, of the i'th species.  Because the individual free energies are replicas of each other, their divergences will be the same as in (\ref{free3}). So when we extract them from the total free energy (\ref{totalfree}), the sum rules (\ref{sumrules}) imply that these terms vanish. Of course, this means that in the regulated theory the divergences reappear as the $m_i \to \infty$ divergences. These terms are renormalized by the counterterms in the effective action. In principle, we would have to compute the finite terms in (\ref{free3}) to identify them. However, here we can use a shortcut, noting that that the counterterms are defined by taking the limits $m_i \to \infty$ {\it at the same rate}. Thus these divergences will behave in exactly the same way as the blueshift divergences which occur when we move the brick wall to the horizon. Since the blueshift formula yieds
$E_{blue} = E/g^{1/2} = E \sqrt{r_H/(2\epsilon)}$, we can simply trade $m_i^2 \ln (\ep/r_H)$ for $m_i^2 \ln(m_i^2)$ in each contribution $\propto m^2_i \ln(\ep/r_H)$ in the sum of $F^i$'s. This yields the dominant contribution to the regulated free energy in the limit $m_i \rightarrow \infty$\footnote{There is also a purely logarithmic divergence coming from the first logarithmic term in (\ref{free3}). We expect that this should match the renormalization of the $({\rm curvature})^2$ couplings as in \cite{Demers:1995dq}, if we extend the Gibbons-Hawking entropy to the Wald entropy, but we will not pursue that here.}
\beq
F= - \frac{N \pi}{12}\ \frac{r_H^3}{\beta^2} \sum_{i=0}^5 \Delta_i m_i^2\log m_i^2+\dots \, .
\eeq

Given the free energy $F(\beta)$, $S = \beta^2 \partial F/\partial \beta$. So the leading divergent contribution to the entropy as the cutoff is taken to infinity is 
\beq
S =  \frac{N \pi}{6} \frac{r_H^3}{\beta} \sum_{i=0}^5 \Delta_i m_i^2\log m_i^2 `\, .
\eeq
This is the leading contribution of $N$ species of particles and their Pauli-Villars regulators to the total entropy in de Sitter causal patch, coming predominantly from the modes which are accumulated near the horizon.
Now, since this system is in equilibrium with the background, we set the temperature $\beta^{-1}$ to the Bekenstein-Gibbons-Hawking temperature of de Sitter space $T_{GH} = H/2\pi$, or alternatively we use $r_H = 1/H = \beta/(2\pi)$. Recalling that the horizon area is $A = 4\pi r_H^2$, 
\beq
S =  \frac{N}{48 \pi} A  \sum_{i=0}^5 \Delta_i m_i^2\log m_i^2 =  \frac{ A}{48 \pi}  \delta~,
\eeq 
where we have employed the definition of the counterterm $\delta$ from 
(\ref{counters}).  When we add this UV  contribution to the bare Bekenstein-Gibbons-Hawking entropy of de Sitter, we obtain simply the finite renormalized entropy
\beq
S^{ren} = S_{dS}+ S  = \frac{A}{4 G_N} + \frac{A}{48 \pi}   \delta = \frac{A}{4 G^{ren}_N} \, .
\eeq
The divergences match: the leading order $N$ dependence precisely cancels. So the species problem never appears when the de Sitter entropy is correctly calculated using the physical renormalized Newton's constant.

We note that this conclusion is expected to remain correct even if the entropy is calculated as the entanglement entropy (from fields with spins $<2$). In the case of black holes, the contribution of background fields to the entanglement entropy and Gibbons-Hawking free energy match precisely.  A similar argument holds for quantum fields on de Sitter backgrounds \cite{Cooperman:2013iqr,Ben-Ami:2015zsa}\footnote{Calculations of the de Sitter entropy in flat slicing can be found in \cite{Muller:1995mz,KeskiVakkuri:2003vj,Maldacena:2012xp}.}. 

\section{Effective field theory and inflation}

The discussion in the previous sections sets the stage for the analysis of inflation in theories with many light matter species. Only after we have renormalized the gravitational sector of the theory, as well as the standard local QFT matter sector, can we consider the question of corrections to the scalar and tensor power spectrum of CMB fluctuations, the inflaton sector dynamics and the viability of long inflation against both perturbative and non-perturbative corrections from both field theory and quantum gravity. 

With these points in mind, we will review the general aspects of large field inflation driven by axions, and the conditions and reasons for its viability as a QFT.   We will then revisit some of the explicit arguments in \cite{Conlon:2012tz} regarding non-perturbative quantum gravity effects,  and reconsider their implications for the correctly renormalized low energy theory.

\subsection{Inflation with many species}

The standard picture of inflation and its main observable prediction, the CMB fluctuations, rely on the validity of semiclassical 4d gravity at the Hubble scale $H$, which is the curvature scale of the background during the inflationary epoch. In light of the discussions above, it is clear that $M^{ren}_{pl}/\sqrt{N} \geq {\cal M}_{UV} \gg H$ is required for this picture to be valid. This is manifest in calculations of loop corrections to the scalar and tensor power spectrum. The metric couples to matter fields with Planck-suppressed couplings, so a 4-d calculation of the density fluctuations (assuming the locally Lorentzian vacuum, aka the Bunch-Davies vacuum, for the inflaton and graviton fluctuations) will produce a spectrum of perturbations \cite{Kaloper:2002uj,Weinberg:2005vy,Senatore:2009cf,Giddings:2010nc}\footnote{Up to logarithmic corrections \cite{Weinberg:2005vy,Senatore:2009cf,Giddings:2010nc}.}
\beq
	P = P_{tree}\left(1 + c N \frac{H^2}{M_{pl}^2} + \ldots \right) =  P_{tree}\left(1 + c' \frac{H^2}{{\cal M}_{UV}^2} + \ldots \right)
\eeq
for both scalar and tensor modes, where $c' \ll 1$ if ${\cal M}_{UV} \ll M_4/\sqrt{N}$. For example, the $N$ species could be Kaluza-Klein modes; the resulting UV scale is the 10d Planck scale \cite{Kaloper:2002uj}. This picture arises from a general effective field theory analysis of the inflaton-graviton sector \cite{Kaloper:2002uj}: higher powers of $H$ come from terms in the effective action that are of higher power in the curvature, are dictated by the graviton wavefunction renormalization, and will be suppressed some UV scale ${\cal M}_{UV}$, which plays the role of a cutoff of the low energy theory.  It is clear that the corrections are only small if ${\cal M}_{UV}  > H$, which is at any rate required for the validity of semiclassical gravity at the scale $H$. For inflation with the minimal required number of efoldings, there is also the question of initial states which deviate from the Bunch-Davies vacuum, which we will ignore in what follows. As long as the inflationary dynamics obeys the standard rules of EFT, these deviations are limited \cite{Porrati:2004gz}. The main physical observables do not depend significantly on the number of light species to the leading order, because they are automatically expressed in terms of the renormalized 4d physical quantities.

\subsection{Axions: inflation and $N$-flation}

\subsubsection{Axions as inflatons}

Axions are perfect candidates for inflatons: the periodicity of the axion $\phi \equiv \phi + 2\pi f_a$ protects the potential from perturbative corrections, allowing for a relatively shallow potential. In the cases that the potential is generated by a dilute gas of instantons and takes the form $V \sim \Lambda^4 \cos (\phi/f_a)$, inflation requires a fairly large value of $f_a \sim M_{pl}$, to support long and uninterrupted slow roll regime that can sustain at least $\sim 60$ efolds of inflation \cite{Freese:1990rb}. There has been much work on constructing axion inflation models, and we will not review that work here. Instead, we will focus on the aspects of axion-driven inflation, with a quasi-de Sitter geometry, relevant for understanding possible entropy bounds. 

As  noted above, to ask questions about the validity of entropy bounds in a quasi-de Sitter space, we must first determine the regime of validity of the semiclassical theory. Ref. \cite{Conlon:2012tz}\ argues that the proper cutoff at which to evaluate the de Sitter entropy in a theory with an axion is the axion decay constant $f_a$.  The argument is that the composite operator $\phi^2(0)$, evaluated with a momentum cutoff ${\cal M}_{UV}$, scales as  ${\cal M}_{UV}^2$.  So, the argument goes, when ${\cal M}_{UV}\sim f_a$, the fluctuations in the scalar field completely delocalize it on the circle $\phi \equiv \phi + 2 \pi f_a$, preventing the semiclassical description of the scalar as a rolling in a single perturbative sector, and smearing it over the full covering space.  
An alternative reading is that the two-point function $\langle \phi(x)\phi(y)\rangle \sim 1/|x-y|^2$ at short distances, and for separations $|x-y| \sim f_a^{-1}$ the fluctuations between points cover the entire target space circle.

As we will now explain, this argument is not correct.  In fact, the cutoff ${\cal M}_{UV}$ can be either larger or small than $f_a$ without leading to any inconsistency.  First of all, if the cutoff  ${\cal M}_{UV}$ is \emph{smaller} than the period $f_a$, fluctuations at the cutoff would obviously do little in the way of smearing the expectation value of the axion over the scale $f_a$. This might be countered by claiming that the cutoff should be close to $M_{pl}$. Yet, as we have seen previously, this is not the case in many models of interest. Secondly, the estimate of the scale of $\phi^2(0)$ in \cite{Conlon:2012tz}\  ignores the renormalization of this composite operator.  In fact the estimate $\langle \phi^2(0) \rangle \sim {\cal M}_{UV}^2$ is really the regularized quadratic divergence of the cosmological constant term in de Sitter space with a massive scalar field, and it will be subtracted off in the correct renormalization procedure. Indeed, in the flat space vacuum, $\langle \phi^2(0) \rangle = 0$ after properly renormalizing the operator.  In de Sitter space, there is an IR contribution only to the renormalized operator, and $\langle \phi^2(0) \rangle \sim H^2$. 

The alternate point that $\langle \phi(x)\phi(y)\rangle \sim 1/|x-y|^2$  leads to the axion being delocalized at the scale $1/f_a$ is true. However, the correct interpretation of this phenomenon is that $f_a$ is the strong coupling scale for the \emph{axion} dynamics.  The axion potential typically arises from instantons which couple to the axion via an irrelevant operator $\sim \frac{\phi}{f_a} F \wedge F$.  Periodicity of the axion guarantees that any direct dependence on $\phi$ (as opposed to its derivatives) must be a periodic function of $\phi/f$, ie a harmonic series in $\cos\phi/f_a$. If we write a low-energy effective field theory by expanding this about a minimum, the expansion will be in powers of $\phi/f_a$, thus indicating that $f_a$ is a natural scale for strong coupling, beyond which the potential cannot be approximated by the first few terms in the expansion.

There is no reason for the axion sector strong coupling scale to be the one at which 4d semiclassical gravity breaks down. For example, for the cases where $f_a < {\cal M}_{UV}$, the UV completion of the axion sector can be described fully in the four-dimensional EFT below ${\cal M}_{UV}$. The axion can be UV-completed as a phase of a Peccei-Quinn complex doublet $\Phi = \varphi e^{i \theta}$, with potential $V(\Phi) = \lambda (|\Phi|^2 - f_a^2)^2$. After symmetry breaking and integrating out the heavy radial mode, the axion decay constant is the Peccei-Quinn {\it vev} $f_a$. The mass of the radial mode is $m_\varphi \sim \sqrt{\lambda} f_a$, and as long as the theory is weakly coupled, $\lambda \ll 1$, we have $m_\varphi \ll f_a$. The cutoff of the low energy theory with only the phase retained is  $\sim m_\varphi$, where the low energy theory of the axion with interactions governed by higher dimension operators, generated after integrating the radial mode, becomes strongly coupled and violates unitarity. To resolve this, all one needs is to integrate the radial mode back in at scales above $m_\varphi$.  This happens entirely within the realm of the EFT with gravity below ${\cal M}_{UV}$. In more complicated cases, as in string theory compactifications, the UV completion will depend on the details of moduli stabilization.  Of course one should understand this UV completion properly when accounting for the axion sector's contribution to the de Sitter entropy below the scale ${\cal M}_{UV}$.

For axion decay constants near the Planck scale, four-dimensional gravity typically breaks down at scales well {\it below} $f_a$, as also noted in \cite{Heidenreich:2015wga}. In particular, if the Kaluza-Klein scale is below $f_a$, most string theory axions  lift in 10 or 11 dimensions to a higher-form gauge field at this scale. The worry arises if one requires $f_a > M_{pl}$ as in the early models of axion inflation \cite{Freese:1990rb}. The QFT sector of such models appears to behave without a problem. However nonperturbative gravity effects -- as exemplified by wormhole calculations of the corrections ot the low energy actions -- may be very dangerous for such models \cite{Giddings:1987cg,Abbott:1989jw,Coleman:1989zu,Kallosh:1995hi,ArkaniHamed:2006dz}. Moreover, the WGC is in tension with elementary axion theories with such large $f_a$ \cite{ArkaniHamed:2006dz,delaFuente:2014aca,Montero:2015ofa,Brown:2015iha,Heidenreich:2015wga,Palti:2015xra}.\footnote{Ref. \cite{Kooner:2015rza}, on the other hand, claims these are not problematic in principle, but that there are problems with using them for inflation in specific string models.}

However it is possible to realize low energy axion models with a very large effective $f_a$, above the actual strong coupling scale of the theory (whether the strong coupling dynamics is from local QFT degrees of freedom, or from quantum gravity). Examples are provided by various realizations of axion monodromy. We provide a specific example here for illustrative purposes, inspired by \cite{Kim:2004rp,Berg:2009tg}. The purpose is not to build a complete model of inflation, but to illustrate how to generate a hierarchy between an effective $f_a$ and the actual strong coupling scale ${\cal M}_{UV} \ll f_a$, within field theory.

Consider a simple case involving two axions, coupling via topological terms to three different gauge groups, in non-orthogonal linear combinations:
\beq
{\cal L}_{int} = \frac{\phi_1}{f_1}  tr F_1 \wedge F_1 +  \frac{\phi_2}{f_2}  tr F_2 \wedge F_2 + \left(\frac{\phi_1}{f_1} - \frac{n\phi_2}{f_2} \right) tr F_3 \wedge F_3 \, ,
\eeq
where $n$ is an integer. 
Provided that all the gauge sectors are weakly coupled just below the cutoff, we can calculate the instanton potential generated by the gauge theories in the dilute gas approximation (when it applies), and find to leading order
\beq
V_{eff} = \mu_1^4 \cos(\frac{\phi_1}{f_1}) + \mu_2^4 \cos(\frac{\phi_2}{f_2}) + \mu_3^4 \cos(\frac{\phi_1}{f_1} - \frac{n\phi_2}{f_2}) \, .
\label{effpot}
\eeq
Let the axion decay constants be comparable, $f_1 \sim f_2 < {\cal M}_{UV}$, but let there be a hierarchy between the gauge sector strong coupling scales $\mu_1 \ll \mu_2 \ll \mu_3$. This can be arranged by a choice of the fermionic charges in the theory, gauge groups, and their coupling constants. 

To understand the perturbative behavior of the theory, pick a particular vacuum of the theory, say  $\phi_1 = \phi_2 = 0$, and consider small fluctuations. The potential (\ref{effpot}) yields the mass matrix of the small fluctuations,
\beq
V_{masses} = \frac{\mu_1^4}{2 f_1^2}  \phi_1^2 +
\frac{\mu_2^4}{2 f_2^2}  \phi_2^2  + 
\frac{n^2 \mu_3^4}{2f_2^2} \left(\phi_2 - \frac{f_2}{n f_1} \phi_1\right)^2
\, .
\label{masses}
\eeq
Given the scale hierarchy, (\ref{masses}) shows that the heaviest field in the theory is really the linear combination $\chi_{heavy} \propto \phi_2 - \frac{f_2}{n f_1} \phi_1$, which is mostly $\phi_2$, with a small admixture of $\phi_1$. So to understand the low energy dynamics, we can pick it as one of the two normal modes of the system,
and choose the direction orthogonal to it as the other. Picking canonical normalizations for these fields yields
\beq
\chi_{heavy} = \frac{f_1 f_2}{f_{eff}} \left( \frac{n \phi_2}{f_2} - \frac{\phi_1}{f_1} \right) \, , ~~~~~~~~~~ 
\chi_{light} =  \frac{f_1 f_2}{f_{eff}} \left( \frac{\phi_2}{f_1} +  \frac{n \phi_1}{f_2} \right)  \, ,
\eeq
where $f_{eff} = \sqrt{n^2 f_1^2 + f_2^2}$. Substituting these field redefinitions into the potential (\ref{effpot}) yields
\beq
V_{eff} = \mu_1^4 \cos(\frac{n \chi_{light} }{f_{eff}} - \frac{f_2\chi_{heavy}}{f_1 f_{eff}} ) + \mu_2^4  \cos(\frac{n f_1 \chi_{heavy} }{f_2 f_{eff}}  + 
\frac{\chi_{light}}{f_{eff}} )  + \mu_3^4 \cos(\frac{f_{eff} \chi_{heavy}}{f_1 f_2}) \, 
\label{effpotnew}
\eeq
with canonically normalized kinetic terms. 

The hierarchy of strong coupling scales means that the last term strongly localizes $\chi_{heavy}$ in a minimum of the final term in (\ref{effpotnew}).  The first term gives a small sinusoidal modulation of the second term, which is a cosine potential for $\chi_{light}$ with periodicity $f_{eff} \sim n f_{1}$.  The trajectory of this field in the coordinates $\phi_1,\phi_2$ can be seen in Figure 2; we have produced a form of axion monodromy.

Let us study the dynamics of $\chi_{light}$ in more detail, and understand when the low-energy effective action becomes strongly coupled. Vacua of the theory correspond to non-zero $\chi_{light}, \chi_{heavy}$: the arguments of the cosines must be odd integer multiples of $\pi$. Specifically, pick a vacuum $\chi_{heavy} = (2l+1) \frac{f_1 f_2}{f_{eff}} \pi$ for the heavy degree of freedom. Expanding the potential (\ref{effpotnew}) about it, and taking $n \gg \mu_2^2/\mu_1^2$ the effective potential for $\chi_{light}$ becomes 
\beq
V_{eff} \simeq \frac{\mu_2^4}{2 f_{eff}^2} \left( \chi_{light} + (2l+1) \frac{n f_1^2}{f_{eff}} \pi \right)^2 + \mu_1^4 \cos\left(\frac{n \chi_{light} }{f_{eff}} - (2l+1) \frac{f_2}{f_{eff}} \pi \right) \, . 
\label{eq:taeffpot}
\eeq
We do not expand the second term: the frequency of this harmonic is much larger than that in the first term, by $n \gg \mu^2_2/\mu^2_1$; and the phase shift is small, $\sim f_2/f_{eff} \sim 1/n$, for many vacua in the theory.  The cosine is merely a harmonic modulation on top of the first term as long as the field $\chi_{light}$ is far away from the vacuum, which is approximately at $\simeq (2l+1)\frac{n f_1^2}{f_{eff}} \pi \simeq (2l+1) f_1$. The potential energy stored in this vacuum displacement is quite small, on the order of 
$V_{eff} \la (2l+1)^2 \mu_2^4$, safely in the regime of semiclassical gravity as long as $\sqrt{2l+1} \mu_2  \la M_{pl}$. 

The field displacement of $\chi_{light}$ from its vacuum can be larger than $M_{pl}$ even when $f_1$ is safely below the Planck scale, if $l \ga M_{pl}/f_1$, and the low energy action (\ref{eq:taeffpot})\ can still remain. This situation is depicted in Fig. 2, where the slanted lines denote the trajectory of the light field $\chi_{light}$, and the field region between different segments belongs to field configurations with energies much larger than those along the slanted lines. 
\begin{figure}[thb]
\label{monodromy}
\centering
\includegraphics[height=7.5cm]{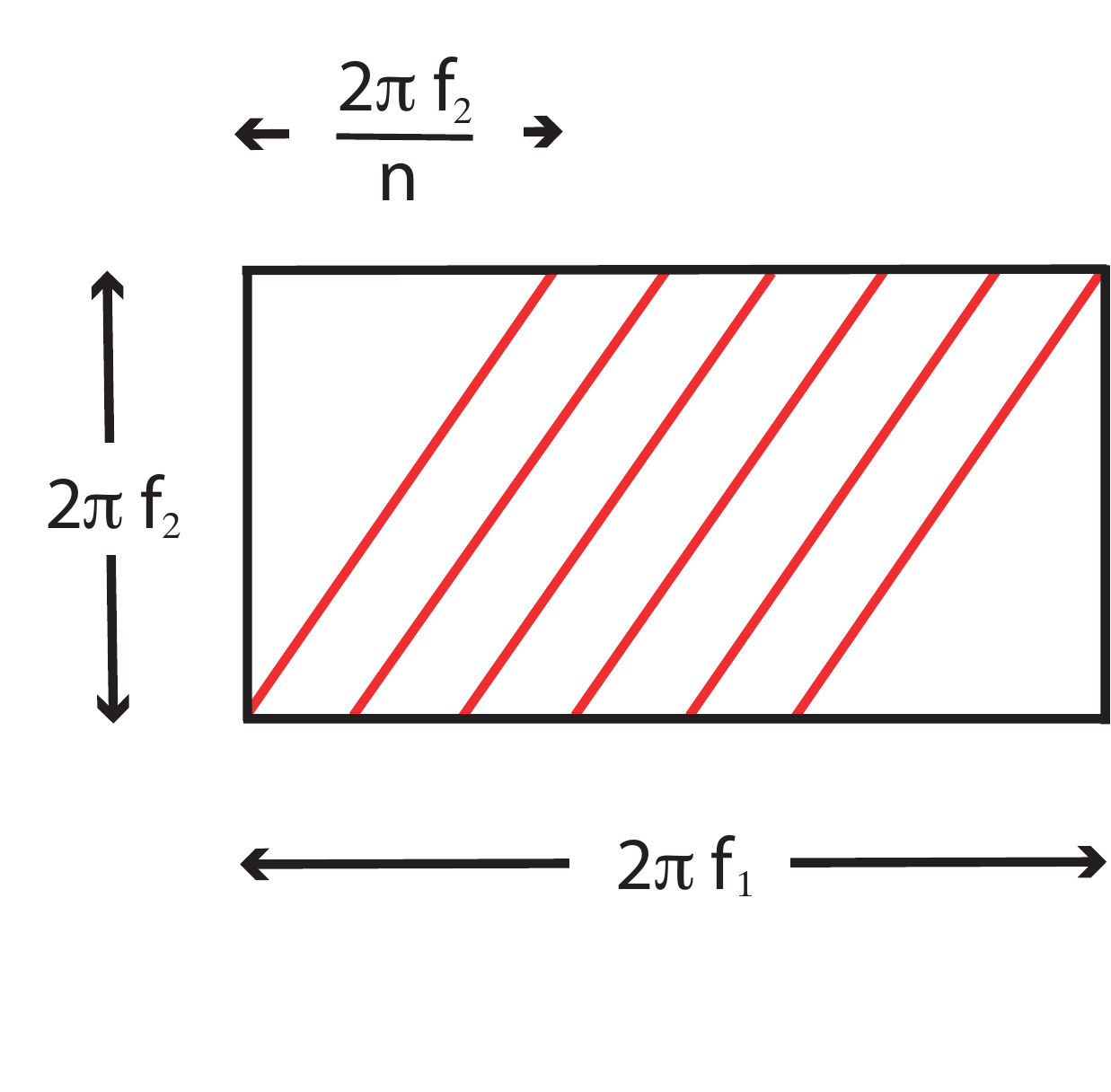}
\caption{A simple axion monodromy from two axions.}
\end{figure}

The cutoff limiting the regime of validity of the single light axion $\chi_{light}$ low energy theory is given by the mass of the heavy field which has been integrated out, ${\cal M}^{eff} \simeq \frac{n \mu_3^2}{f_2} \sim \frac{n^2 \mu_3^2}{f_{eff}}$. This can be easily arranged to be smaller than the apparent strong coupling scale $f_{eff}$ of the dynamics for $\chi_{light}$, by choosing $\mu_3 \la f_{eff}/n \sim f_1$. Thus the theory can be fully constructed in the regime where the standard perturbation theory operates, with low energy shift symmetry protecting the flatness of the effective potential, and the phase misalignment between the two axions simulating the transplanckian field displacements required for supporting long inflation. The nonperturbative corrections from gauge theory, and even wormhole induced terms from quantum gravity remain small. Finally, in light of the discussion beforehand, when the theory is correctly renormalized it also automatically obeys the covariant entropy bounds. More general monodromy models work in a similar way.

\subsubsection{$N$-flation}

$N$-flation \cite{Dimopoulos:2005ac}\ proposes to achieve effectively super-Planckian inflaton range from a large number of axions with sub-Planckian decay constants. The underlying assumption is that there are many axions which are displaced from their potential minima and are light. The total energy that drives inflation comes from the sum of the individual energies for each field, and this is what supports the slow roll of each individual field. So for $N$ axions rolling in unison, the effective inflaton range can scale as $\sqrt{N} f_a$, where $f_a$ is a characteristic fundamental axion decay constant and $N$ is the number of axions. 

There is an active, ongoing discussion in the literature as to whether such a theory can be embedded in a good string theory model, and whether it is consistent with a properly interpreted version of the Weak Gravity conjecture: see for example \cite{Brown:2015iha,Brown:2015lia,Bachlechner:2014gfa,Bachlechner:2015qja,Rudelius:2014wla,Rudelius:2015xta,Heidenreich:2015wga,Heidenreich:2015nta}. This is an interesting  question, and we will not address it here.  Our point is merely that there is no obvious violation of the covariant entropy bounds in this example.  Ref. \cite{Dimopoulos:2005ac}\ already noted that a large number of species can run in loops and correct the bare value of Newton's constant.  As argued above, 4d semiclassical gravity will break down at some scale ${\cal M}_{UV} \leq M_4/\sqrt{N}$, and there is no intrinsic problem with the covariant entropy bound.  

The real issue is the value of ${\cal M}_{UV}$. In addition to requiring ${\cal M}_{UV}> H$, there is also an open question for the calculability of the underlying string compactification if ${\cal M}_{UV}$ is lower than all of the compactification scale, the string scale, and the 10d Planck scale.  In this case, the 4d dynamics relevant for the compactification intrinsically requires understanding strongly coupled quantum gravity; this could, for example, complicate considerations of moduli stabilization.  

\subsection{Flux compactifications with large gauge groups}

Ref. \cite{Conlon:2012tz} examines the one-K\"ahler-modulus KKLT model \cite{Kachru:2003aw}\ and its generalization to racetrack models  \cite{BlancoPillado:2004ns}, when the nonperturbative potential for the Kahler modulus is generated by gaugino condensation on wrapped D7-branes, and claims that the species problem renders them inconsistent. As we have argued, it does not.  However it is interesting to ask at what scale 4d semiclassical gravity is expected to break down, and what dynamics becomes relevant.  

Let us consider the the scenario described in \cite{Kachru:2003aw}. The modulus $\sigma = (R_{KK}/\ell_{p,10})^4$ is the volume modulus; the gauge coupling is then $g_{YM}^2 = 4\pi/\sigma$.  The superpotential generated by gaugino condensation is taken to be
\beq
	W = W_0 + A e^{- 2\pi \sigma/N_7}
\eeq
where $N_7$ is the rank of the D7-brane gauge group, and $W_0$ is the tree-level flux-induced term in the superpotential.  If, following \cite{Kachru:2003aw,Conlon:2012tz}, we take $W_0 \ll 1$, in order to get a solution well-described by classical 10d geometry. In this case, the Kahler modulus is parametrically
\beq
	\sigma = \left(\frac{R_{KK}}{\ell_{p,10}}\right)^4 \sim \frac{N_7}{2\pi} |\ln W_0| \, .
\eeq
Now using the fact that $m_{pl,4}^2 \sim R_{KK}^6/\ell_{p,10}^8$, and taking the number of light species to be $\sim N_7^2$ (more precisely, we are assuming here that $N_7$ is the dominant contribution to this number), we find
\beq
	{\cal M}_{UV}^2 = \frac{m_{p,4}^2}{N_7^2}  \sim \frac{|\ln W_0|^2}{(2\pi R_{KK})^2}
\eeq
in other words, for small $W_0$, the strong coupling scale is at or larger than the Kaluza-Klein scale of the string theory compactification.  Thus, before (or when) this scale is reached, 4d semiclassical gravity has already broken down in a completely standard fashion.

While the potentials in \cite{BlancoPillado:2004ns} are more complicated, we may take the specific numbers used in eqs. (3.11-3.13) of that paper and find once again that the stromg coupling scale $\Lambda$ is of of the same order as the Kaluza-klein scale. This is preserved under the rescalings of parameters in sec. 3.3 of that paper.

\vskip.8cm
	
{\bf Acknowledgments}: 
Part of this work was carried out while A.L. was at the Aspen Center for Physics, which is supported by National Science Foundation grant PHY-1066293.  A.L. would like to thank the organizers of and participants in the "Primordial Physics" workshop for stimulating conversations on subjects related to this work. NK would like to thank CP$^3$-Origins, University of Southern Denmark, and CCPP, NYU, for kind hospitality in the course of this work.
We would like to thank A. Arvanitaki, A. Brown, G. D'Amico, M. Porrati, and G. Zahariade for useful discussions. N.K. is supported in part by the DOE Grant DE-SC0009999. A.L. is supported by the DOE grant DE-SC0009987. M.S.S. is supported by the Lundbeck Foundation.  MK is supported in part by the NSF through grant PHY-1214302, and he acknowledges membership at the NYU-ECNU Joint Physics Research Institute in Shanghai.

\bibliographystyle{utphys}
\bibliography{axmod_refs.bib}

\providecommand{\href}[2]{#2}\begingroup\raggedright\begin{thebibliography}{100}

\bibitem{Lyth:1996im}
D.~H. Lyth, ``{What would we learn by detecting a gravitational wave signal in
  the cosmic microwave background anisotropy?},''
  \href{http://dx.doi.org/10.1103/PhysRevLett.78.1861}{{\em Phys. Rev. Lett.}
  {\bfseries 78} (1997) 1861--1863},
\href{http://arxiv.org/abs/hep-ph/9606387}{{\ttfamily arXiv:hep-ph/9606387}}.

\bibitem{Efstathiou:2005tq}
G.~Efstathiou and K.~J. Mack, ``{The Lyth Bound Revisited},''
  \href{http://dx.doi.org/10.1088/1475-7516/2005/05/008}{{\em JCAP} {\bfseries
  0505} (2005) 008},
\href{http://arxiv.org/abs/astro-ph/0503360}{{\ttfamily
  arXiv:astro-ph/0503360}}.

\bibitem{ArkaniHamed:2006dz}
N.~Arkani-Hamed, L.~Motl, A.~Nicolis, and C.~Vafa, ``{The string landscape,
  black holes and gravity as the weakest force},'' {\em JHEP} {\bfseries 06}
  (2007) 060,
\href{http://arxiv.org/abs/hep-th/0601001}{{\ttfamily arXiv:hep-th/0601001}}.

\bibitem{Bousso:1999xy}
R.~Bousso, ``{A Covariant entropy conjecture},''
  \href{http://dx.doi.org/10.1088/1126-6708/1999/07/004}{{\em JHEP} {\bfseries
  07} (1999) 004},
\href{http://arxiv.org/abs/hep-th/9905177}{{\ttfamily arXiv:hep-th/9905177
  [hep-th]}}.

\bibitem{Freese:1990rb}
K.~Freese, J.~A. Frieman, and A.~V. Olinto, ``{Natural inflation with pseudo -
  Nambu-Goldstone bosons},''
\href{http://dx.doi.org/10.1103/PhysRevLett.65.3233}{{\em Phys. Rev. Lett.}
  {\bfseries 65} (1990) 3233--3236}.

\bibitem{ArkaniHamed:2003mz}
N.~Arkani-Hamed, H.-C. Cheng, P.~Creminelli, and L.~Randall, ``{Pseudonatural
  inflation},'' \href{http://dx.doi.org/10.1088/1475-7516/2003/07/003}{{\em
  JCAP} {\bfseries 0307} (2003) 003},
\href{http://arxiv.org/abs/hep-th/0302034}{{\ttfamily arXiv:hep-th/0302034}}.

\bibitem{ArkaniHamed:2003wu}
N.~Arkani-Hamed, H.-C. Cheng, P.~Creminelli, and L.~Randall, ``{Extranatural
  inflation},'' \href{http://dx.doi.org/10.1103/PhysRevLett.90.221302}{{\em
  Phys. Rev. Lett.} {\bfseries 90} (2003) 221302},
\href{http://arxiv.org/abs/hep-th/0301218}{{\ttfamily arXiv:hep-th/0301218}}.

\bibitem{Dimopoulos:2005ac}
S.~Dimopoulos, S.~Kachru, J.~McGreevy, and J.~G. Wacker, ``{N-flation},''
  \href{http://dx.doi.org/10.1088/1475-7516/2008/08/003}{{\em JCAP} {\bfseries
  0808} (2008) 003},
\href{http://arxiv.org/abs/hep-th/0507205}{{\ttfamily arXiv:hep-th/0507205}}.

\bibitem{Brown:2015iha}
J.~Brown, W.~Cottrell, G.~Shiu, and P.~Soler, ``{Fencing in the Swampland:
  Quantum Gravity Constraints on Large Field Inflation},''
  \href{http://dx.doi.org/10.1007/JHEP10(2015)023}{{\em JHEP} {\bfseries 10}
  (2015) 023},
\href{http://arxiv.org/abs/1503.04783}{{\ttfamily arXiv:1503.04783 [hep-th]}}.

\bibitem{Heidenreich:2015wga}
B.~Heidenreich, M.~Reece, and T.~Rudelius, ``{Weak Gravity Strongly Constrains
  Large-Field Axion Inflation},''
\href{http://arxiv.org/abs/1506.03447}{{\ttfamily arXiv:1506.03447 [hep-th]}}.

\bibitem{Palti:2015xra}
E.~Palti, ``{On Natural Inflation and Moduli Stabilisation in String Theory},''
  \href{http://dx.doi.org/10.1007/JHEP10(2015)188}{{\em JHEP} {\bfseries 10}
  (2015) 188},
\href{http://arxiv.org/abs/1508.00009}{{\ttfamily arXiv:1508.00009 [hep-th]}}.

\bibitem{Silverstein:2008sg}
E.~Silverstein and A.~Westphal, ``{Monodromy in the CMB: Gravity Waves and
  String Inflation},'' \href{http://dx.doi.org/10.1103/PhysRevD.78.106003}{{\em
  Phys. Rev.} {\bfseries D78} (2008) 106003},
\href{http://arxiv.org/abs/0803.3085}{{\ttfamily arXiv:0803.3085 [hep-th]}}.

\bibitem{McAllister:2008hb}
L.~McAllister, E.~Silverstein, and A.~Westphal, ``{Gravity Waves and Linear
  Inflation from Axion Monodromy},''
  \href{http://dx.doi.org/10.1103/PhysRevD.82.046003}{{\em Phys.Rev.}
  {\bfseries D82} (2010) 046003},
\href{http://arxiv.org/abs/0808.0706}{{\ttfamily arXiv:0808.0706 [hep-th]}}.

\bibitem{Kaloper:2008fb}
N.~Kaloper and L.~Sorbo, ``{A Natural Framework for Chaotic Inflation},''
  \href{http://dx.doi.org/10.1103/PhysRevLett.102.121301}{{\em Phys. Rev.
  Lett.} {\bfseries 102} (2009) 121301},
\href{http://arxiv.org/abs/0811.1989}{{\ttfamily arXiv:0811.1989 [hep-th]}}.

\bibitem{Kaloper:2011jz}
N.~Kaloper, A.~Lawrence, and L.~Sorbo, ``{An Ignoble Approach to Large Field
  Inflation},'' \href{http://dx.doi.org/10.1088/1475-7516/2011/03/023}{{\em
  JCAP} {\bfseries 1103} (2011) 023},
\href{http://arxiv.org/abs/1101.0026}{{\ttfamily arXiv:1101.0026 [hep-th]}}.

\bibitem{DAmico:2012ji}
G.~D'Amico, R.~Gobbetti, M.~Kleban, and M.~Schillo, ``{Unwinding Inflation},''
  \href{http://dx.doi.org/10.1088/1475-7516/2013/03/004}{{\em JCAP} {\bfseries
  1303} (2013) 004},
\href{http://arxiv.org/abs/1211.4589}{{\ttfamily arXiv:1211.4589 [hep-th]}}.

\bibitem{DAmico:2012sz}
G.~D'Amico, R.~Gobbetti, M.~Schillo, and M.~Kleban, ``{Inflation from Flux
  Cascades},'' \href{http://dx.doi.org/10.1016/j.physletb.2013.07.050}{{\em
  Phys.Lett.} {\bfseries B725} (2013) 218--222},
\href{http://arxiv.org/abs/1211.3416}{{\ttfamily arXiv:1211.3416 [hep-th]}}.

\bibitem{DAmico:2013iaa}
G.~D'Amico, R.~Gobbetti, M.~Kleban, and M.~Schillo, ``{Large-scale anomalies
  from primordial dissipation},''
  \href{http://dx.doi.org/10.1088/1475-7516/2013/11/013}{{\em JCAP} {\bfseries
  1311} (2013) 013},
\href{http://arxiv.org/abs/1306.6872}{{\ttfamily arXiv:1306.6872
  [astro-ph.CO]}}.

\bibitem{Palti:2014kza}
E.~Palti and T.~Weigand, ``{Towards large r from [p,q]-inflation},''
\href{http://arxiv.org/abs/1403.7507}{{\ttfamily arXiv:1403.7507 [hep-th]}}.

\bibitem{Hebecker:2014eua}
A.~Hebecker, S.~C. Kraus, and L.~T. Witkowski, ``{D7-Brane Chaotic
  Inflation},''
\href{http://arxiv.org/abs/1404.3711}{{\ttfamily arXiv:1404.3711 [hep-th]}}.

\bibitem{Ibanez:2014kia}
L.~E. Ibanez and I.~Valenzuela, ``{The Inflaton as a MSSM Higgs and Open String
  Modulus Monodromy Inflation},''
\href{http://arxiv.org/abs/1404.5235}{{\ttfamily arXiv:1404.5235 [hep-th]}}.

\bibitem{Ibanez:2014swa}
L.~E. Ibanez, F.~Marchesano, and I.~Valenzuela, ``{Higgs-otic Inflation and
  String Theory},'' \href{http://dx.doi.org/10.1007/JHEP01(2015)128}{{\em JHEP}
  {\bfseries 01} (2015) 128},
\href{http://arxiv.org/abs/1411.5380}{{\ttfamily arXiv:1411.5380 [hep-th]}}.

\bibitem{Bielleman:2015lka}
S.~Bielleman, L.~E. Ibanez, F.~G. Pedro, and I.~Valenzuela, ``{Multifield
  Dynamics in Higgs-Otic Inflation},''
\href{http://arxiv.org/abs/1505.00221}{{\ttfamily arXiv:1505.00221 [hep-th]}}.

\bibitem{Bielleman:2015ina}
S.~Bielleman, L.~E. Ibanez, and I.~Valenzuela, ``{Minkowski 3-forms, Flux
  String Vacua, Axion Stability and Naturalness},''
\href{http://arxiv.org/abs/1507.06793}{{\ttfamily arXiv:1507.06793 [hep-th]}}.

\bibitem{Grimm:2014vva}
T.~W. Grimm, ``{Axion Inflation in F-theory},''
\href{http://arxiv.org/abs/1404.4268}{{\ttfamily arXiv:1404.4268 [hep-th]}}.

\bibitem{Blumenhagen:2014gta}
R.~Blumenhagen and E.~Plauschinn, ``{Towards Universal Axion Inflation and
  Reheating in String Theory},''
\href{http://arxiv.org/abs/1404.3542}{{\ttfamily arXiv:1404.3542 [hep-th]}}.

\bibitem{Blumenhagen:2014nba}
R.~Blumenhagen, D.~Herschmann, and E.~Plauschinn, ``{The Challenge of Realizing
  F-term Axion Monodromy Inflation in String Theory},''
  \href{http://dx.doi.org/10.1007/JHEP01(2015)007}{{\em JHEP} {\bfseries 01}
  (2015) 007},
\href{http://arxiv.org/abs/1409.7075}{{\ttfamily arXiv:1409.7075 [hep-th]}}.

\bibitem{Marchesano:2014mla}
F.~Marchesano, G.~Shiu, and A.~M. Uranga, ``{F-term Axion Monodromy
  Inflation},''
\href{http://arxiv.org/abs/1404.3040}{{\ttfamily arXiv:1404.3040 [hep-th]}}.

\bibitem{Gao:2014uha}
X.~Gao, T.~Li, and P.~Shukla, ``{Combining Universal and Odd RR Axions for
  Aligned Natural Inflation},''
  \href{http://dx.doi.org/10.1088/1475-7516/2014/10/048}{{\em JCAP} {\bfseries
  1410} (2014) 048},
\href{http://arxiv.org/abs/1406.0341}{{\ttfamily arXiv:1406.0341 [hep-th]}}.

\bibitem{Conlon:2012tz}
J.~P. Conlon, ``{Quantum Gravity Constraints on Inflation},''
  \href{http://dx.doi.org/10.1088/1475-7516/2012/09/019}{{\em JCAP} {\bfseries
  1209} (2012) 019},
\href{http://arxiv.org/abs/1203.5476}{{\ttfamily arXiv:1203.5476 [hep-th]}}.

\bibitem{Banks:2003pt}
T.~Banks and W.~Fischler, ``{An Upper bound on the number of e-foldings},''
\href{http://arxiv.org/abs/astro-ph/0307459}{{\ttfamily arXiv:astro-ph/0307459
  [astro-ph]}}.

\bibitem{Kaloper:2004gp}
N.~Kaloper, M.~Kleban, and L.~Sorbo, ``{Observational implications of
  cosmological event horizons},''
  \href{http://dx.doi.org/10.1016/j.physletb.2004.08.068}{{\em Phys. Lett.}
  {\bfseries B600} (2004) 7--14},
\href{http://arxiv.org/abs/astro-ph/0406099}{{\ttfamily arXiv:astro-ph/0406099
  [astro-ph]}}.

\bibitem{Stelle:1976gc}
K.~S. Stelle, ``{Renormalization of Higher Derivative Quantum Gravity},''
\href{http://dx.doi.org/10.1103/PhysRevD.16.953}{{\em Phys. Rev.} {\bfseries
  D16} (1977) 953--969}.

\bibitem{Kim:2004rp}
J.~E. Kim, H.~P. Nilles, and M.~Peloso, ``{Completing natural inflation},''
  \href{http://dx.doi.org/10.1088/1475-7516/2005/01/005}{{\em JCAP} {\bfseries
  0501} (2005) 005},
\href{http://arxiv.org/abs/hep-ph/0409138}{{\ttfamily arXiv:hep-ph/0409138
  [hep-ph]}}.

\bibitem{Berg:2009tg}
M.~Berg, E.~Pajer, and S.~Sjors, ``{Dante's Inferno},''
  \href{http://dx.doi.org/10.1103/PhysRevD.81.103535}{{\em Phys.Rev.}
  {\bfseries D81} (2010) 103535},
\href{http://arxiv.org/abs/0912.1341}{{\ttfamily arXiv:0912.1341 [hep-th]}}.

\bibitem{Kabat:1995eq}
D.~N. Kabat, ``{Black hole entropy and entropy of entanglement},''
  \href{http://dx.doi.org/10.1016/0550-3213(95)00443-V}{{\em Nucl. Phys.}
  {\bfseries B453} (1995) 281--299},
\href{http://arxiv.org/abs/hep-th/9503016}{{\ttfamily arXiv:hep-th/9503016
  [hep-th]}}.

\bibitem{Larsen:1995ax}
F.~Larsen and F.~Wilczek, ``{Renormalization of black hole entropy and of the
  gravitational coupling constant},''
  \href{http://dx.doi.org/10.1016/0550-3213(95)00548-X}{{\em Nucl. Phys.}
  {\bfseries B458} (1996) 249--266},
\href{http://arxiv.org/abs/hep-th/9506066}{{\ttfamily arXiv:hep-th/9506066
  [hep-th]}}.

\bibitem{Callan:1994py}
C.~G. Callan, Jr. and F.~Wilczek, ``{On geometric entropy},''
  \href{http://dx.doi.org/10.1016/0370-2693(94)91007-3}{{\em Phys. Lett.}
  {\bfseries B333} (1994) 55--61},
\href{http://arxiv.org/abs/hep-th/9401072}{{\ttfamily arXiv:hep-th/9401072
  [hep-th]}}.

\bibitem{Jacobson:2012ek}
T.~Jacobson and A.~Satz, ``{Black hole entanglement entropy and the
  renormalization group},''
  \href{http://dx.doi.org/10.1103/PhysRevD.87.084047}{{\em Phys. Rev.}
  {\bfseries D87} no.~8, (2013) 084047},
\href{http://arxiv.org/abs/1212.6824}{{\ttfamily arXiv:1212.6824}}.

\bibitem{Cooperman:2013iqr}
J.~H. Cooperman and M.~A. Luty, ``{Renormalization of Entanglement Entropy and
  the Gravitational Effective Action},''
  \href{http://dx.doi.org/10.1007/JHEP12(2014)045}{{\em JHEP} {\bfseries 12}
  (2014) 045},
\href{http://arxiv.org/abs/1302.1878}{{\ttfamily arXiv:1302.1878 [hep-th]}}.

\bibitem{Donnelly:2011hn}
W.~Donnelly, ``{Decomposition of entanglement entropy in lattice gauge
  theory},'' \href{http://dx.doi.org/10.1103/PhysRevD.85.085004}{{\em Phys.
  Rev.} {\bfseries D85} (2012) 085004},
\href{http://arxiv.org/abs/1109.0036}{{\ttfamily arXiv:1109.0036 [hep-th]}}.

\bibitem{Casini:2013rba}
H.~Casini, M.~Huerta, and J.~A. Rosabal, ``{Remarks on entanglement entropy for
  gauge fields},'' \href{http://dx.doi.org/10.1103/PhysRevD.89.085012}{{\em
  Phys. Rev.} {\bfseries D89} no.~8, (2014) 085012},
\href{http://arxiv.org/abs/1312.1183}{{\ttfamily arXiv:1312.1183 [hep-th]}}.

\bibitem{Huang:2014pfa}
K.-W. Huang, ``{Central Charge and Entangled Gauge Fields},''
  \href{http://dx.doi.org/10.1103/PhysRevD.92.025010}{{\em Phys. Rev.}
  {\bfseries D92} no.~2, (2015) 025010},
\href{http://arxiv.org/abs/1412.2730}{{\ttfamily arXiv:1412.2730 [hep-th]}}.

\bibitem{Ghosh:2015iwa}
S.~Ghosh, R.~M. Soni, and S.~P. Trivedi, ``{On The Entanglement Entropy For
  Gauge Theories},'' \href{http://dx.doi.org/10.1007/JHEP09(2015)069}{{\em
  JHEP} {\bfseries 09} (2015) 069},
\href{http://arxiv.org/abs/1501.02593}{{\ttfamily arXiv:1501.02593 [hep-th]}}.

\bibitem{Solodukhin:2015hma}
S.~N. Solodukhin, ``{Newton constant, contact terms and entropy},''
  \href{http://dx.doi.org/10.1103/PhysRevD.91.084028}{{\em Phys. Rev.}
  {\bfseries D91} no.~8, (2015) 084028},
\href{http://arxiv.org/abs/1502.03758}{{\ttfamily arXiv:1502.03758 [hep-th]}}.

\bibitem{Donnelly:2015hxa}
W.~Donnelly and A.~C. Wall, ``{Geometric entropy and edge modes of the
  electromagnetic field},''
\href{http://arxiv.org/abs/1506.05792}{{\ttfamily arXiv:1506.05792 [hep-th]}}.

\bibitem{Soni:2015yga}
R.~M. Soni and S.~P. Trivedi, ``{Aspects of Entanglement Entropy for Gauge
  Theories},''
\href{http://arxiv.org/abs/1510.07455}{{\ttfamily arXiv:1510.07455 [hep-th]}}.

\bibitem{Ma:2015xes}
C.-T. Ma, ``{Entanglement with Centers},''
\href{http://arxiv.org/abs/1511.02671}{{\ttfamily arXiv:1511.02671 [hep-th]}}.

\bibitem{tHooft:1974bx}
G.~'t~Hooft and M.~J.~G. Veltman, ``{One loop divergencies in the theory of
  gravitation},''
{\em Annales Poincare Phys. Theor.} {\bfseries A20} (1974) 69--94.

\bibitem{Adler:1982ri}
S.~L. Adler, ``{Einstein Gravity as a Symmetry Breaking Effect in Quantum Field
  Theory},'' \href{http://dx.doi.org/10.1103/RevModPhys.54.729}{{\em Rev. Mod.
  Phys.} {\bfseries 54} (1982) 729}.
[Erratum: Rev. Mod. Phys.55,837(1983)].

\bibitem{Fursaev:1994ea}
D.~V. Fursaev and S.~N. Solodukhin, ``{On one loop renormalization of black
  hole entropy},'' \href{http://dx.doi.org/10.1016/0370-2693(95)01290-7}{{\em
  Phys. Lett.} {\bfseries B365} (1996) 51--55},
\href{http://arxiv.org/abs/hep-th/9412020}{{\ttfamily arXiv:hep-th/9412020
  [hep-th]}}.

\bibitem{Demers:1995dq}
J.-G. Demers, R.~Lafrance, and R.~C. Myers, ``{Black hole entropy without brick
  walls},'' \href{http://dx.doi.org/10.1103/PhysRevD.52.2245}{{\em Phys. Rev.}
  {\bfseries D52} (1995) 2245--2253},
\href{http://arxiv.org/abs/gr-qc/9503003}{{\ttfamily arXiv:gr-qc/9503003
  [gr-qc]}}.

\bibitem{Kaloper:2013zca}
N.~Kaloper and A.~Padilla, ``{Sequestering the Standard Model Vacuum Energy},''
  \href{http://dx.doi.org/10.1103/PhysRevLett.112.091304}{{\em Phys. Rev.
  Lett.} {\bfseries 112} no.~9, (2014) 091304},
\href{http://arxiv.org/abs/1309.6562}{{\ttfamily arXiv:1309.6562 [hep-th]}}.

\bibitem{Kaloper:2014dqa}
N.~Kaloper and A.~Padilla, ``{Vacuum Energy Sequestering: The Framework and Its
  Cosmological Consequences},''
  \href{http://dx.doi.org/10.1103/PhysRevD.90.084023,
  10.1103/PhysRevD.90.109901}{{\em Phys. Rev.} {\bfseries D90} no.~8, (2014)
  084023}, \href{http://arxiv.org/abs/1406.0711}{{\ttfamily arXiv:1406.0711
  [hep-th]}}.
[Addendum: Phys. Rev.D90,no.10,109901(2014)].

\bibitem{Kaloper:2015jra}
N.~Kaloper, A.~Padilla, D.~Stefanyszyn, and G.~Zahariade, ``{A Manifestly Local
  Theory of Vacuum Energy Sequestering},''
\href{http://arxiv.org/abs/1505.01492}{{\ttfamily arXiv:1505.01492 [hep-th]}}.

\bibitem{Veneziano:2001ah}
G.~Veneziano, ``{Large N bounds on, and compositeness limit of, gauge and
  gravitational interactions},''
  \href{http://dx.doi.org/10.1088/1126-6708/2002/06/051}{{\em JHEP} {\bfseries
  06} (2002) 051},
\href{http://arxiv.org/abs/hep-th/0110129}{{\ttfamily arXiv:hep-th/0110129
  [hep-th]}}.

\bibitem{Dvali:2007wp}
G.~Dvali and M.~Redi, ``{Black Hole Bound on the Number of Species and Quantum
  Gravity at LHC},'' \href{http://dx.doi.org/10.1103/PhysRevD.77.045027}{{\em
  Phys. Rev.} {\bfseries D77} (2008) 045027},
\href{http://arxiv.org/abs/0710.4344}{{\ttfamily arXiv:0710.4344 [hep-th]}}.

\bibitem{Kabat:1994vj}
D.~N. Kabat and M.~J. Strassler, ``{A Comment on entropy and area},''
  \href{http://dx.doi.org/10.1016/0370-2693(94)90515-0}{{\em Phys. Lett.}
  {\bfseries B329} (1994) 46--52},
\href{http://arxiv.org/abs/hep-th/9401125}{{\ttfamily arXiv:hep-th/9401125
  [hep-th]}}.

\bibitem{Solodukhin:1995ak}
S.~N. Solodukhin, ``{One loop renormalization of black hole entropy due to
  nonminimally coupled matter},''
  \href{http://dx.doi.org/10.1103/PhysRevD.52.7046}{{\em Phys. Rev.} {\bfseries
  D52} (1995) 7046--7052},
\href{http://arxiv.org/abs/hep-th/9504022}{{\ttfamily arXiv:hep-th/9504022
  [hep-th]}}.

\bibitem{Simon:1990ic}
J.~Z. Simon, ``{Higher Derivative Lagrangians, Nonlocality, Problems and
  Solutions},''
\href{http://dx.doi.org/10.1103/PhysRevD.41.3720}{{\em Phys. Rev.} {\bfseries
  D41} (1990) 3720}.

\bibitem{Simon:1990jn}
J.~Z. Simon, ``{The Stability of flat space, semiclassical gravity, and higher
  derivatives},''
\href{http://dx.doi.org/10.1103/PhysRevD.43.3308}{{\em Phys. Rev.} {\bfseries
  D43} (1991) 3308--3316}.

\bibitem{Simon:1991bm}
J.~Z. Simon, ``{No Starobinsky inflation from selfconsistent semiclassical
  gravity},''
\href{http://dx.doi.org/10.1103/PhysRevD.45.1953}{{\em Phys. Rev.} {\bfseries
  D45} (1992) 1953--1960}.

\bibitem{Randall:1999vf}
L.~Randall and R.~Sundrum, ``{An alternative to compactification},''
  \href{http://dx.doi.org/10.1103/PhysRevLett.83.4690}{{\em Phys. Rev. Lett.}
  {\bfseries 83} (1999) 4690--4693},
\href{http://arxiv.org/abs/hep-th/9906064}{{\ttfamily arXiv:hep-th/9906064}}.

\bibitem{Gubser:1999vj}
S.~S. Gubser, ``{AdS / CFT and gravity},''
  \href{http://dx.doi.org/10.1103/PhysRevD.63.084017}{{\em Phys. Rev.}
  {\bfseries D63} (2001) 084017},
\href{http://arxiv.org/abs/hep-th/9912001}{{\ttfamily arXiv:hep-th/9912001
  [hep-th]}}.

\bibitem{Sakharov:1975zg}
A.~D. Sakharov, ``{Spectral Density of Eigenvalues of the Wave Equation and the
  Vacuum Polarization},'' \href{http://dx.doi.org/10.1007/BF01036152}{{\em
  Theor. Math. Phys.} {\bfseries 23} (1976) 435--444}.
[Sov. Phys. Usp.34,395(1991)].

\bibitem{Susskind:1993ws}
L.~Susskind, ``{Some speculations about black hole entropy in string theory},''
\href{http://arxiv.org/abs/hep-th/9309145}{{\ttfamily arXiv:hep-th/9309145
  [hep-th]}}.

\bibitem{Solodukhin:1994yz}
S.~N. Solodukhin, ``{The Conical singularity and quantum corrections to entropy
  of black hole},'' \href{http://dx.doi.org/10.1103/PhysRevD.51.609}{{\em Phys.
  Rev.} {\bfseries D51} (1995) 609--617},
\href{http://arxiv.org/abs/hep-th/9407001}{{\ttfamily arXiv:hep-th/9407001
  [hep-th]}}.

\bibitem{Dvali:2008jb}
G.~Dvali and S.~N. Solodukhin, ``{Black Hole Entropy and Gravity Cutoff},''
\href{http://arxiv.org/abs/0806.3976}{{\ttfamily arXiv:0806.3976 [hep-th]}}.

\bibitem{Susskind:1993if}
L.~Susskind, L.~Thorlacius, and J.~Uglum, ``{The Stretched horizon and black
  hole complementarity},''
  \href{http://dx.doi.org/10.1103/PhysRevD.48.3743}{{\em Phys. Rev.} {\bfseries
  D48} (1993) 3743--3761},
\href{http://arxiv.org/abs/hep-th/9306069}{{\ttfamily arXiv:hep-th/9306069
  [hep-th]}}.

\bibitem{Jacobson:1994iw}
T.~Jacobson, ``{Black hole entropy and induced gravity},''
\href{http://arxiv.org/abs/gr-qc/9404039}{{\ttfamily arXiv:gr-qc/9404039
  [gr-qc]}}.

\bibitem{Hawking:2000da}
S.~Hawking, J.~M. Maldacena, and A.~Strominger, ``{de Sitter entropy, quantum
  entanglement and AdS / CFT},''
  \href{http://dx.doi.org/10.1088/1126-6708/2001/05/001}{{\em JHEP} {\bfseries
  05} (2001) 001},
\href{http://arxiv.org/abs/hep-th/0002145}{{\ttfamily arXiv:hep-th/0002145
  [hep-th]}}.

\bibitem{Emparan:2006ni}
R.~Emparan, ``{Black hole entropy as entanglement entropy: A Holographic
  derivation},'' \href{http://dx.doi.org/10.1088/1126-6708/2006/06/012}{{\em
  JHEP} {\bfseries 06} (2006) 012},
\href{http://arxiv.org/abs/hep-th/0603081}{{\ttfamily arXiv:hep-th/0603081
  [hep-th]}}.

\bibitem{Kaloper:2012hu}
N.~Kaloper, ``{Cutoffs, Stretched Horizons and Black Hole Radiators},''
  \href{http://dx.doi.org/10.1103/PhysRevD.86.104052}{{\em Phys. Rev.}
  {\bfseries D86} (2012) 104052},
\href{http://arxiv.org/abs/1203.3455}{{\ttfamily arXiv:1203.3455 [hep-th]}}.

\bibitem{Kim:1998zs}
W.~T. Kim, ``{Entropy of (2+1)-dimensional de Sitter space in terms of brick
  wall method},'' \href{http://dx.doi.org/10.1103/PhysRevD.59.047503}{{\em
  Phys. Rev.} {\bfseries D59} (1999) 047503},
\href{http://arxiv.org/abs/hep-th/9810169}{{\ttfamily arXiv:hep-th/9810169
  [hep-th]}}.

\bibitem{Strominger:1996sh}
A.~Strominger and C.~Vafa, ``{Microscopic origin of the Bekenstein-Hawking
  entropy},'' \href{http://dx.doi.org/10.1016/0370-2693(96)00345-0}{{\em Phys.
  Lett.} {\bfseries B379} (1996) 99--104},
\href{http://arxiv.org/abs/hep-th/9601029}{{\ttfamily arXiv:hep-th/9601029
  [hep-th]}}.

\bibitem{PhysRevA.43.2046}
J.~M. Deutsch, ``Quantum statistical mechanics in a closed system,''
  \href{http://dx.doi.org/10.1103/PhysRevA.43.2046}{{\em Phys. Rev. A}
  {\bfseries 43} (Feb, 1991) 2046--2049}.

\bibitem{PhysRevE.50.888}
M.~Srednicki, ``Chaos and quantum thermalization,''
  \href{http://dx.doi.org/10.1103/PhysRevE.50.888}{{\em Phys. Rev. E}
  {\bfseries 50} (Aug, 1994) 888--901}.

\bibitem{PhysRevLett.80.1373}
H.~Tasaki, ``From quantum dynamics to the canonical distribution: General
  picture and a rigorous example,''
  \href{http://dx.doi.org/10.1103/PhysRevLett.80.1373}{{\em Phys. Rev. Lett.}
  {\bfseries 80} (Feb, 1998) 1373--1376}.

\bibitem{rigol2008thermalization}
M.~Rigol, V.~Dunjko, and M.~Olshanii, ``Thermalization and its mechanism for
  generic isolated quantum systems,'' {\em Nature} {\bfseries 452} no.~7189,
  (2008) 854--858.

\bibitem{Dvali:2007hz}
G.~Dvali, ``{Black Holes and Large N Species Solution to the Hierarchy
  Problem},'' \href{http://dx.doi.org/10.1002/prop.201000009}{{\em Fortsch.
  Phys.} {\bfseries 58} (2010) 528--536},
\href{http://arxiv.org/abs/0706.2050}{{\ttfamily arXiv:0706.2050 [hep-th]}}.

\bibitem{Randall:1999ee}
L.~Randall and R.~Sundrum, ``{A large mass hierarchy from a small extra
  dimension},'' \href{http://dx.doi.org/10.1103/PhysRevLett.83.3370}{{\em Phys.
  Rev. Lett.} {\bfseries 83} (1999) 3370--3373},
\href{http://arxiv.org/abs/hep-ph/9905221}{{\ttfamily arXiv:hep-ph/9905221}}.

\bibitem{Verlinde:1999fy}
H.~L. Verlinde, ``{Holography and compactification},''
  \href{http://dx.doi.org/10.1016/S0550-3213(00)00224-8}{{\em Nucl. Phys.}
  {\bfseries B580} (2000) 264--274},
\href{http://arxiv.org/abs/hep-th/9906182}{{\ttfamily arXiv:hep-th/9906182
  [hep-th]}}.

\bibitem{tHooft:1984re}
G.~'t~Hooft, ``{On the Quantum Structure of a Black Hole},''
\href{http://dx.doi.org/10.1016/0550-3213(85)90418-3}{{\em Nucl. Phys.}
  {\bfseries B256} (1985) 727}.

\bibitem{Ben-Ami:2015zsa}
O.~Ben-Ami, D.~Carmi, and M.~Smolkin, ``{Renormalization group flow of
  entanglement entropy on spheres},''
  \href{http://dx.doi.org/10.1007/JHEP08(2015)048}{{\em JHEP} {\bfseries 08}
  (2015) 048},
\href{http://arxiv.org/abs/1504.00913}{{\ttfamily arXiv:1504.00913 [hep-th]}}.

\bibitem{Muller:1995mz}
R.~Muller and C.~O. Lousto, ``{Entanglement entropy in curved space-times with
  event horizons},'' \href{http://dx.doi.org/10.1103/PhysRevD.52.4512}{{\em
  Phys. Rev.} {\bfseries D52} (1995) 4512--4517},
\href{http://arxiv.org/abs/gr-qc/9504049}{{\ttfamily arXiv:gr-qc/9504049
  [gr-qc]}}.

\bibitem{KeskiVakkuri:2003vj}
E.~Keski-Vakkuri and M.~S. Sloth, ``{Holographic bounds on the UV cutoff scale
  in inflationary cosmology},''
  \href{http://dx.doi.org/10.1088/1475-7516/2003/08/001}{{\em JCAP} {\bfseries
  0308} (2003) 001},
\href{http://arxiv.org/abs/hep-th/0306070}{{\ttfamily arXiv:hep-th/0306070
  [hep-th]}}.

\bibitem{Maldacena:2012xp}
J.~Maldacena and G.~L. Pimentel, ``{Entanglement entropy in de Sitter space},''
  \href{http://dx.doi.org/10.1007/JHEP02(2013)038}{{\em JHEP} {\bfseries 02}
  (2013) 038},
\href{http://arxiv.org/abs/1210.7244}{{\ttfamily arXiv:1210.7244 [hep-th]}}.

\bibitem{Kaloper:2002uj}
N.~Kaloper, M.~Kleban, A.~E. Lawrence, and S.~Shenker, ``{Signatures of short
  distance physics in the cosmic microwave background},''
  \href{http://dx.doi.org/10.1103/PhysRevD.66.123510}{{\em Phys. Rev.}
  {\bfseries D66} (2002) 123510},
\href{http://arxiv.org/abs/hep-th/0201158}{{\ttfamily arXiv:hep-th/0201158}}.

\bibitem{Weinberg:2005vy}
S.~Weinberg, ``{Quantum contributions to cosmological correlations},''
  \href{http://dx.doi.org/10.1103/PhysRevD.72.043514}{{\em Phys. Rev.}
  {\bfseries D72} (2005) 043514},
\href{http://arxiv.org/abs/hep-th/0506236}{{\ttfamily arXiv:hep-th/0506236
  [hep-th]}}.

\bibitem{Senatore:2009cf}
L.~Senatore and M.~Zaldarriaga, ``{On Loops in Inflation},''
  \href{http://dx.doi.org/10.1007/JHEP12(2010)008}{{\em JHEP} {\bfseries 12}
  (2010) 008},
\href{http://arxiv.org/abs/0912.2734}{{\ttfamily arXiv:0912.2734 [hep-th]}}.

\bibitem{Giddings:2010nc}
S.~B. Giddings and M.~S. Sloth, ``{Semiclassical relations and IR effects in de
  Sitter and slow-roll space-times},''
  \href{http://dx.doi.org/10.1088/1475-7516/2011/01/023}{{\em JCAP} {\bfseries
  1101} (2011) 023},
\href{http://arxiv.org/abs/1005.1056}{{\ttfamily arXiv:1005.1056 [hep-th]}}.

\bibitem{Porrati:2004gz}
M.~Porrati, ``{Bounds on generic high-energy physics modifications to the
  primordial power spectrum from back reaction on the metric},''
  \href{http://dx.doi.org/10.1016/j.physletb.2004.06.090}{{\em Phys. Lett.}
  {\bfseries B596} (2004) 306--310},
\href{http://arxiv.org/abs/hep-th/0402038}{{\ttfamily arXiv:hep-th/0402038
  [hep-th]}}.

\bibitem{Giddings:1987cg}
S.~B. Giddings and A.~Strominger, ``{Axion Induced Topology Change in Quantum
  Gravity and String Theory},''
\href{http://dx.doi.org/10.1016/0550-3213(88)90446-4}{{\em Nucl. Phys.}
  {\bfseries B306} (1988) 890}.

\bibitem{Abbott:1989jw}
L.~F. Abbott and M.~B. Wise, ``{Wormholes and global symmetries},''
\href{http://dx.doi.org/10.1016/0550-3213(89)90503-8}{{\em Nucl. Phys.}
  {\bfseries B325} (1989) 687}.

\bibitem{Coleman:1989zu}
S.~R. Coleman and K.-M. Lee, ``{Wormholes Made Without Massless Matter
  Fields},''
\href{http://dx.doi.org/10.1016/0550-3213(90)90149-8}{{\em Nucl. Phys.}
  {\bfseries B329} (1990) 387}.

\bibitem{Kallosh:1995hi}
R.~Kallosh, A.~D. Linde, D.~A. Linde, and L.~Susskind, ``{Gravity and global
  symmetries},'' \href{http://dx.doi.org/10.1103/PhysRevD.52.912}{{\em Phys.
  Rev.} {\bfseries D52} (1995) 912--935},
\href{http://arxiv.org/abs/hep-th/9502069}{{\ttfamily arXiv:hep-th/9502069}}.

\bibitem{delaFuente:2014aca}
A.~de~la Fuente, P.~Saraswat, and R.~Sundrum, ``{Natural Inflation and Quantum
  Gravity},'' \href{http://dx.doi.org/10.1103/PhysRevLett.114.151303}{{\em
  Phys. Rev. Lett.} {\bfseries 114} no.~15, (2015) 151303},
\href{http://arxiv.org/abs/1412.3457}{{\ttfamily arXiv:1412.3457 [hep-th]}}.

\bibitem{Montero:2015ofa}
M.~Montero, A.~M. Uranga, and I.~Valenzuela, ``{Transplanckian axions!?},''
  \href{http://dx.doi.org/10.1007/JHEP08(2015)032}{{\em JHEP} {\bfseries 08}
  (2015) 032},
\href{http://arxiv.org/abs/1503.03886}{{\ttfamily arXiv:1503.03886 [hep-th]}}.

\bibitem{Kooner:2015rza}
K.~Kooner, S.~Parameswaran, and I.~Zavala, ``{Warping the Weak Gravity
  Conjecture},''
\href{http://arxiv.org/abs/1509.07049}{{\ttfamily arXiv:1509.07049 [hep-th]}}.

\bibitem{Brown:2015lia}
J.~Brown, W.~Cottrell, G.~Shiu, and P.~Soler, ``{On Axionic Field Ranges,
  Loopholes and the Weak Gravity Conjecture},''
\href{http://arxiv.org/abs/1504.00659}{{\ttfamily arXiv:1504.00659 [hep-th]}}.

\bibitem{Bachlechner:2014gfa}
T.~C. Bachlechner, C.~Long, and L.~McAllister, ``{Planckian Axions in String
  Theory},''
\href{http://arxiv.org/abs/1412.1093}{{\ttfamily arXiv:1412.1093 [hep-th]}}.

\bibitem{Bachlechner:2015qja}
T.~C. Bachlechner, C.~Long, and L.~McAllister, ``{Planckian Axions and the Weak
  Gravity Conjecture},''
\href{http://arxiv.org/abs/1503.07853}{{\ttfamily arXiv:1503.07853 [hep-th]}}.

\bibitem{Rudelius:2014wla}
T.~Rudelius, ``{On the Possibility of Large Axion Moduli Spaces},''
  \href{http://dx.doi.org/10.1088/1475-7516/2015/04/049}{{\em JCAP} {\bfseries
  1504} no.~04, (2015) 049},
\href{http://arxiv.org/abs/1409.5793}{{\ttfamily arXiv:1409.5793 [hep-th]}}.

\bibitem{Rudelius:2015xta}
T.~Rudelius, ``{Constraints on Axion Inflation from the Weak Gravity
  Conjecture},'' \href{http://dx.doi.org/10.1088/1475-7516/2015/09/020,
  10.1088/1475-7516/2015/9/020}{{\em JCAP} {\bfseries 1509} no.~09, (2015)
  020},
\href{http://arxiv.org/abs/1503.00795}{{\ttfamily arXiv:1503.00795 [hep-th]}}.

\bibitem{Heidenreich:2015nta}
B.~Heidenreich, M.~Reece, and T.~Rudelius, ``{Sharpening the Weak Gravity
  Conjecture with Dimensional Reduction},''
\href{http://arxiv.org/abs/1509.06374}{{\ttfamily arXiv:1509.06374 [hep-th]}}.

\bibitem{Kachru:2003aw}
S.~Kachru, R.~Kallosh, A.~D. Linde, and S.~P. Trivedi, ``{De Sitter vacua in
  string theory},'' \href{http://dx.doi.org/10.1103/PhysRevD.68.046005}{{\em
  Phys. Rev.} {\bfseries D68} (2003) 046005},
\href{http://arxiv.org/abs/hep-th/0301240}{{\ttfamily arXiv:hep-th/0301240
  [hep-th]}}.

\bibitem{BlancoPillado:2004ns}
J.~J. Blanco-Pillado, C.~P. Burgess, J.~M. Cline, C.~Escoda, M.~Gomez-Reino,
  R.~Kallosh, A.~D. Linde, and F.~Quevedo, ``{Racetrack inflation},''
  \href{http://dx.doi.org/10.1088/1126-6708/2004/11/063}{{\em JHEP} {\bfseries
  11} (2004) 063},
\href{http://arxiv.org/abs/hep-th/0406230}{{\ttfamily arXiv:hep-th/0406230
  [hep-th]}}.

\end{thebibliography}\endgroup

\end{document}